\documentclass[12pt]{article}
\usepackage{amsfonts,amsmath,amstext,amsthm,latexsym,graphicx,bezier}
\newtheorem{theorem}{Theorem}[section]

\begin{document}
\begin{titlepage}
%\newpage
\begin{center}
{\bf \large Direct and Inverse Computational Methods for Electromagnetic Scattering in Biological Diagnostics}\vspace{1.5cm}\\
\end{center}

%\begin{center}
%A synopsis submitted in partial fulfilment of the
%requirements for admission to a doctorate degree in mathematics at M$\ddot{ \rm a}$lardalen university, Sweden\vspace{1.5cm}\\
%\end{center}
%{\bf \underline{Proposed supervisors}}\vspace{0.3cm}\\
%{\bf Sweden} ~~~~~~~~~~~~~~~~~~~~~~~~~~~~~~~~~~~~~~~~~~~~~~~~~~~~~~~~~{\bf Local supervisor}\\

Farid Monsefi\\
School of Education, Culture and Communication (UKK),\\
Department of Innovation, Design, and Technique (IDT),\\
M$\ddot{ \rm a}$lardalen University, Sweden
\vspace*{0.6cm}

Magnus Otterskog\\
Department of Innovation, Design, and Technique (IDT),\\
M$\ddot{ \rm a}$lardalen University, Sweden\\
\vspace*{0.2cm}

Sergei Silvestrov\\
%Professor of Mathematics/Applied Mathematics,
%Subject Chair for Mathematics/Applied Mathematics,
School of Education, Culture and Communication (UKK),\\
M$\ddot{ \rm a}$lardalen University, Sweden\\
\vspace*{0.6cm}
\begin{center}
November 2013
\end{center}

\end{titlepage}
\begin{titlepage}
\begin{abstract}
   Scattering theory has had a major roll in twentieth century mathematical physics. Mathematical modeling and algorithms of direct,- and inverse electromagnetic scattering formulation due to biological tissues are investigated. The algorithms are used for a model based illustration technique within the microwave range. A number of methods is given to solve the inverse electromagnetic scattering problem in which the nonlinear and ill-posed nature of the problem are acknowledged.\\*[3ex]
\end{abstract}
\end{titlepage}
\section{Introduction}
Inverse formulations are solved on a daily basis in many disciplines such as image and signal processing, astrophysics, acoustics, quantum mechanics, geophysics and electromagnetic scattering. The inverse formulation, as an interdisciplinary field, involves people from different fields within natural science. To find out the contents of a given black box without opening it, would be a good analogy to describe the general inverse problem. Experiments will be carried on to guess and realize the inner properties of the box. It is common to call the contents of the box "the model" and the result of the experiment "the data". The experiment itself is called "the forward modeling". As sufficient information cannot be provided by an experiment, a process of regularization will be needed. The reason to this issue is that there can be more than one model ('different black boxes') that would produce the same data. On the other hand, improperly posed numerical computations will occur in the calculation procedure. Thus, a process of regularization constitutes a major step to solve the inverse problem. Regularization is used at the moment when selection of the most reasonable model is on focus. Computational methods and techniques ought to be as flexible as possible from case to case. A computational technique utilized for small problems may fail totally when it is used to large numerical domains within the inverse formulation. Hence, new methodologies and algorithms would be created for new problems though existing methods are insufficient. This is the major character of the existing inverse formulation in problems with huge numerical domains. There are both old and new computational tools and techniques for solving linear and nonlinear inverse problems. Linear algebra has been extensively used within linear and nonlinear inverse theory to estimate noise and efficient inverting of large and full matrices. As existing numerical algorithms may fail, new algorithms must be developed to carry out nonlinear inverse problems.\\*[3ex]
Electromagnetic inverse,- and direct scattering problems are, like other related areas, of equal interest. The electromagnetic scattering theory is about the effect an inhomogeneous medium has on an incident wave where the total electromagnetic field is consisted of the incident,- and the scattered field. The direct problem in such context is to determine the scattered field from the knowledge of the incident field and also from the governing wave equation deduced from the Maxwell's equations. As the direct scattering problem has been thoroughly investigated, the inverse scattering problem has not yet a rigorous mathematical/numerical basis. Because the nonlinearity nature of the inverse scattering problem, one will face improperly posed numerical computation. This means that, in particular applications, small perturbations in the measured data cause large errors in the reconstruction of the scatterer. Some regularization methods must be used to remedy the ill-conditioning due to the resulting matrix equations. Concerning the existence of a solution to the inverse electromagnetic scattering one has to think about finding approximate solutions after making the inverse problem stabilized. A number of methods is given to solve the inverse electromagnetic scattering problem in which the nonlinear and ill-posed nature of the problem are acknowledged. Earlier attempts to stabilize the inverse problem was via reducing the problem into a linear integral equation of the first kind. However, general techniques were introduced to treat the inverse problems without applying any integral equation formulation of the problem.
\subsection{Background}
Scattering theory has had a major roll in twentieth century mathematical physics. In computational electromagnetics, the direct scattering problem is to determine a scattered field from knowledge of an incident field and the differential equation governing the wave equation. The incident field is emitted from a source, an antenna for instance, against an inhomogeneous medium. The total field is assumed to be the sum of the incident field and the scattered field. The governing differential equation in such cases is the coupled differential form of Maxwell's equations, which will be converted to the wave equation.\\*[3ex]
In order to guarantee operability of advanced electronic devices and systems, electromagnetic measurements should be compared to results from computational methods. The experimental techniques are expensive and time consuming but are still widely used. Hence, the advantage of obtaining data from tests can be weighted against the large amount of time and expense required to operate such tests. Analytic solution of Maxwell's equations offers many advantages over experimental methods but applicability of analytical electromagnetic modeling is often limited to simple geometries and boundary conditions. As the analytical solutions of Maxwell's equations by the method of \emph{separation of variables} and \emph{series expansions} have a limited scope, they are not applicable in a general case and in a real-world application. Availability of high performance computers during the last decades has been one of the reasons to use numerical techniques within computational modeling to solve Maxwell's equations also for complicated geometries and boundaries.\\*[3ex]
The main objective of this article is to investigate mathematical modeling and algorithms to solve the direct, and inverse electromagnetic scattering problem due to biological tissues for a model based illustration technique within the microwave range. Such algorithms are used to make it possible for parallel processing of the heavy and large numerical calculation due to the inverse formulation of the problem. The parallelism of the calculations can then be performed on GPU:s, CPU:s, and FPGA:s. By the aid of a deeper mathematical analysis and thereby faster numerical algorithms an improvement of the existing numerical algorithms will occur. The algorithms may be in the the time domain, frequency domain and a combination of both domains. %The new algorithms will be developed and implemented in MATLAB to be re-programmed and executed in other proper and more optimized programming languages; this re-programming is needed to apply and suit the algorithms to GPU:s, CPU:s, and FPGA:s.
\clearpage
\section{Related Concepts in Electromagnetism}\footnote{The following two chapters are based, to a large extent, on the work presented in \cite{Farid:Lic.thesis}.}
In constructing the electrostatic model, the electric field intensity
vector $\textbf{E}$ and the electric flux density vector,
$\textbf{D}$, are respectively defined. The fundamental governing
differential equations are \cite{book:Sadiku}
\begin{eqnarray} \label{eq:ma1}
\mathbf{\nabla}\times \mathbf{E} &=&0 \\
\mathbf{\nabla}\cdot \mathbf{D} &=&\rho _{v} \nonumber
\end{eqnarray}
where $\rho _{v}$ is the volume charge density. By introducing $\epsilon=\epsilon_{r}\epsilon_{0}$ as the
the \emph{electric permittivity} where $\epsilon_{r}$ is \emph{relative permittivity}, and $\epsilon_{0}=8.854\times10^{-12} F/m$ as the \emph{permittivity of free space} for a linear and isotropic media, $\textbf{E}$
and $\textbf{D}$ are related by relation
\begin{equation}
\mathbf{D}=\epsilon \mathbf{E}
\end{equation}
The fundamental governing equations for magnetostatic model are
\begin{eqnarray} \label{eq:ma2}
\mathbf{\nabla}\cdot \mathbf{B} &=&0 \\
\mathbf{\nabla}\times \mathbf{H} &=&\mathbf{J} \nonumber
\end{eqnarray}
where $\textbf{B}$ and $\textbf{H}$ are defined as the magnetic flux
density vector and the magnetic field intensity vector, respectively.
$\textbf{B}$ and $\textbf{H}$ are related as
\begin{equation}
\mathbf{H}=\frac{1}{\mu_{0}\mu_{r} }\mathbf{B}
\end{equation}
where $\mu_{0}\mu_{r}=\mu$ is defined as magnetic permeability of the medium which
is measured in $H/m.$; $\mu_{0}=4\pi\times10^{-7}$ $H/m$ is called \emph{permeability of free space} and $\mu_{r}$ is a (material-dependent) number. The medium in question is assumed to be linear
and isotropic. Eqns. (\ref{eq:ma1}) and (\ref{eq:ma2}) are known as
Maxwell's equations and form the foundation of electromagnetic
theory. As it is seen in the above relations, $\textbf{E}$ and
$\textbf{D}$ in the electrostatic model are not related to
$\textbf{B}$ and $\textbf{H}$ in the magnetostatic model. The
coexistence of static electric fields and magnetic electric fields
in a conducting medium causes an electromagnetostatic field and a
time-varying magnetic field gives rise to an electric field. These
are verified by numerous experiments. Static models are not suitable
for explaining time-varying electromagnetic phenomenon. Under
time-varying conditions it is necessary to construct an
electromagnetic model in which the electric field vectors
$\textbf{E}$ and $\textbf{D}$ are related to the magnetic field
vectors $\textbf{B}$ and $\textbf{H}$. In such situations, the
equivalent equations are constructed as
\begin{equation}
\mathbf{\nabla}\times \mathbf{E} = -\frac{\partial
\mathbf{B}}{\partial t}
\end{equation}
\begin{equation}
\mathbf{\nabla}\times \mathbf{H} = \mathbf{J} + \frac{\partial
\mathbf{D}}{\partial t}
\end{equation}
\begin{equation}
\mathbf{\nabla}\cdot \mathbf{D} =\rho _{v}
\end{equation}
\begin{equation}
\mathbf{\nabla}\cdot \mathbf{B} =0
\end{equation}
where $\textbf{J}$ is current density. As it is seen, the Maxwell's
equations above are in differential form. To explain electromagnetic
phenomena in a physical environment, it is more convenient to
convert the differential forms into their integral-form equivalents.
There are several techniques to convert differential equations into
integral equations but in the above cases, one may apply Stokes's
theorem to obtain integral form of Maxwell's equations after taking
the surface integral of both sides of the equations over an open
surface $S$ with contour $C.$ The result will be constructed as in
the following table.%\clearpage
\begin{center}
\textit{Maxwell's equations}\\
\vspace{0.63cm} \hspace{-1.3cm}\textit{Differential form}
\hspace{3cm} \textit{Integral form} \vspace{-0.2cm}
\end{center}
\begin{equation}
\mathbf{\nabla} \times \mathbf{H} = \mathbf{J} + \frac{\partial
\mathbf{D}}{\partial t} \hspace{2cm} \oint \nolimits_C \mathbf{H}
\cdot d\mathbf{L} = I + \int_S \frac{\partial \mathbf{D}}{\partial
t} \cdot d\mathbf{S}
\end{equation}
\begin{equation}
\hspace{-0.2cm}\mathbf{\nabla} \times \mathbf{E} = - \frac{\partial
\mathbf{B}}{\partial t} \hspace{2.6cm} \oint\nolimits_C \mathbf{E}
\cdot d\mathbf{L} = -\int\nolimits_S \frac{\partial
\mathbf{B}}{\partial t}\cdot d\mathbf{S}
\end{equation}
\begin{equation}
\mathbf{\nabla} \cdot \mathbf{D} = \rho_v \hspace{3cm} \int_S
\mathbf{D} \cdot d\mathbf{S} = \int_V \rho_{v} dV
\end{equation}
\begin{equation}
\hspace{-1mm} \mathbf{\nabla} \cdot \mathbf{B} = 0\hspace{3.2cm}
\int_S \mathbf{B} \cdot d\mathbf{S} = 0
\end{equation}
$\rho_v$, in the above table, is the electric charge density in
$C/m^{3}$. %The PEEC method uses the integral form of the Maxwell's
%equations to solve the electromagnetic field quantities and also the
%partial elements.
\subsection{Green's Functions}\normalsize
When a physical system is subject to some external disturbance, a
non-homogeneity arises in the mathematical formulation of the
problem, either in the differential equation or in the auxiliary
conditions or both. When the differential equation is
nonhomogeneous, a particular solution of the equation can be found
by applying either the method of undetermined coefficients or the
variation of parameter technique. In general, however, such
techniques lead to a particular solution that has no special
physical significance. Green's functions\footnote{George Green,
1793-1841, was one of the most remarkable of nineteenth century
physicists, a self-taught mathematician whose work has contributed
greatly to modern physics.} are specific functions that develop
general solution formulas for solving nonhomogeneous differential
equations. Importantly, this type of formulation gives an increased
physical knowledge since every Green's function has a physical
significance. This function measures the response of a system due to
a point source somewhere on the fundamental domain, and all other
solutions due to different source terms are found to be
superpositions \footnote{Consider a set of functions $\phi _{n}$ for
$n=1,2,...,N$. If each number of the functions $\phi _{n}$ is a
solution to the partial differential equation $L\phi =0,$ with $L$
as a linear operator and with some prescribed boundary conditions,
then the linear combination $\phi
_{N}=\phi_{0}+\sum\limits_{n=1}^{N}a_{n}\phi _{n}$ also satisfies
$L\phi =g.$ Here, $g$ is a known excitation or source. This
fundamental concept is verified in different mathematical
literature.} of the Green's function. There are, however, cases
where Green's functions fail to exist, depending on boundaries.
Although Green's first interest was in electrostatics, Green's
mathematics is nearly all devised to solve general physical problems. The
inverse-square law had recently been established
experimentally, and George Green wanted to calculate how this
determined the distribution of charge on the surfaces of conductors.
He made great use of the electrical potential and gave it that name.
Actually, one of the theorems that he proved in this context became
famous and is nowadays known as Green's theorem. It relates the
properties of mathematical functions at the surfaces of a closed
volume to other properties inside. The powerful method of Green's
functions involves what are now called Green's functions, $G(x,x
')$. Applying Green's function method, solution of the
differential equation $Ly=F(x)$, by $L$ as a linear differential
operator, can be written as
\begin{equation}
y(x)=\int\limits_{0}^{x}G(x,x')F(x')dx'.
\end{equation}
To see this, consider the equation
\[
\frac{dy}{dx}+ky=F(x),
\]
which can be solved by the standard integrating factor technique to
give
\[
y=e^{-kx}\int\limits_{0}^{x}e^{kx'}F(x')dx' =
\int\limits_{0}^{x}e^{-k(x-x')}F(x')dx'
\]
so that $G(x,x') = e^{-k(x-x')}$.
This technique may be applied to other more complicated systems. In
an electrical circuit the Green's function is the current due to an
applied voltage pulse. In electrostatics, the Green's function is
the potential due to a change applied at a particular point in
space. In general the Green's function is, as mentioned earlier, the
response of a system to a stimulus applied at a particular point in
space or time. This concept has been readily adapted to quantum
physics where the applied stimulus is the injection of a quantum of
energy.
Within electromagnetic computation, it is common
practice to use two methods for determining the Green's function in
the cases where there is some kind of symmetry in the geometry of
the electromagnetic problem. These are the eigenvalue formulation
and the method of images. These two methods are described in the
following sections, but in order to its importance, the method of
the eigenfunction expansion method is first presented.
\subsection{Green's Functions and Eigenfunctions}
If the eigenvalue problem associated with the operator $L$ can be
solved, then one may find the associated Green's function. It is
known that the eigenvalue problem
\begin{equation} \label{eq:Lulu}
Lu=\lambda u, \hspace{3mm} a<x<b
\end{equation}
by prescribed boundary conditions, has infinite many eigenvalues
and corresponding orthonormal eigenfunctions as $\lambda _{n}$ and
$\phi _{n},$ respectively, where $n=1,2,3,...$ Moreover, the
eigenfunctions form a basis for the square integrable functions on
the interval $(a,b)$. Therefore it is assumed that the solution $u$
is given in terms of eigenfunctions as
\begin{equation} \label{eq:ux}
u(x)=\sum\limits_{n=1}^{\infty }c_{n}\phi _{n}(x)
\end{equation}
where the coefficients $c_{n}$ are to be determined. Further, the
given function $f$ forms the source term in the nonhomogeneous
differential equation
\begin{equation} \label{eq:Luf}
Lu = f \hspace{7mm} or \hspace{7mm} u = L^{-1} f \nonumber
\end{equation}
where $L^{-1}$ is the inverse operator to the operator $L$. Now,
the given function $f$ can be written in terms of the eigenfunctions
as
\begin{equation} \label{eq:fx}
f(x)=\sum\limits_{n=1}^{\infty }f_{n}\phi _{n}(x),
\end{equation}
with
\begin{equation}\label{eq:fx4}
f_{n}=\int\limits_{a}^{b}f(\xi )\phi _{n}(\xi)d\xi.
\end{equation}
Combining (\ref{eq:ux}), (\ref{eq:Luf}), and (\ref{eq:fx4}) gives
\begin{equation}
L\left( \sum\limits_{n=1}^{\infty }c_{n}\phi _{n}(x)\right)
=\sum\limits_{n=1}^{\infty }f_{n}\phi _{n}(x)
\end{equation}
By the linear property associated with superposition principle, it
can be shown that
\begin{equation}
L\left( \sum\limits_{n=1}^{\infty }c_{n}\phi _{n}(x)\right)
=\sum\limits_{n=1}^{\infty }c_{n}L(\phi _{n}(x)).
\end{equation}
But
\begin{equation} \label{eq:cnL}
\sum\limits_{n=1}^{\infty }c_{n}L(\phi
_{n}(x))=\sum\limits_{n=1}^{\infty }c_{n}\lambda _{n}\phi
_{n}(x)=\sum\limits_{n=1}^{\infty }f_{n}\phi _{n}(x),
\end{equation}
which finally yields
\begin{equation} \label{eq:L}
L\left( \sum\limits_{n=1}^{\infty }c_{n}\phi _{n}(x)\right)
=\sum\limits_{n=1}^{\infty }f_{n}\phi _{n}(x).
\end{equation}
By comparing the above equations, it will be obtained that
\begin{equation}
c_{n}=\frac{1}{\lambda _{n}} \textnormal{ and }
f_{n}=\frac{1}{\lambda _{n}} \int \limits_{a}^{b}f(\xi )\phi
_{n}(\xi )d\xi \textnormal{ for } n=1,2,3,...
\end{equation}
Further
\begin{eqnarray} \label{eq:gf1}
u(x) &=&\sum\limits_{n=1}^{\infty }c_{n}\phi _{n}(x) \\
&=&\sum\limits_{n=1}^{\infty }\frac{1}{\lambda _{n}}\left(
\int\limits_{a}^{b}f(\xi )\phi _{n}(\xi )d\xi \right) \phi _{n}(x).
\nonumber
\end{eqnarray}
Now, it is supposed that an interchange of summation and integral is
allowed. In this case (\ref{eq:gf1}) can be written as
\begin{equation}
u(x)=\int\limits_{a}^{b}\left( \sum\limits_{n=1}^{\infty }\frac{\phi
_{n}(x)\phi _{n}(\xi )}{\lambda _{n}}\right) f(\xi )d\xi.
\end{equation}
On the other hand, by the definition of Green's function, one may write
\begin{equation}
u(x)=L^{-1}f=\int\limits_{a}^{b}g(x,\xi )f(\xi )d\xi.
\end{equation}
By comparing the last two equations, $u(x)$ can be expressed in
terms of Green's functions as
\begin{equation}
g(x,\xi )=\sum\limits_{n=1}^{\infty }\frac{\phi _{n}(x)\phi _{n}(\xi )}{%
\lambda _{n}}.
\end{equation}
$g(x,\xi )$ is the Green's function associated with the eigenvalue problem
(\ref{eq:Lulu}) with the differential operator $L$.
\subsection{The Method of Images}\normalsize
Solution of electromagnetic fields is greatly supported and
facilitated by mathematical theorems in vector analysis. Maxwell's
equations are based on Helmholtz's theorem where it is verified that
a vector is uniquely specified by giving its divergence and curl,
within a simply connected region and its normal component over the
boundary. This can be proved as a mathematical theorem in a general
manner \cite{Arfken:book}. Solving partial differential equations
(PDE) like Maxwell's equation desires different methods, depending
on, for instance, which boundary condition the PDE has and in which
physical field it is studied.
The Green's function modeling is an
applicable method to solve Maxwell's equations for some frequently
used cases by different boundary conditions. The issue in this type
of formulation is, in the first hand, determining and solving the
appropriate Green's function by its boundary condition. Once the
Green's function is determined, one may receive a clue to the
physical interpretation of the whole problem and hence a better
understanding of it. This forms the general manner of applying
Green's function formulation in different fields of science. In some
cases within electromagnetic modeling, where the physical source is
in the vicinity of a perfect electric conducting (PEC) surface and
where there is some kind of symmetry in the geometry of the problem,
the method of images will be a logical and facilitating method to
determine the appropriate Green's function. The method of images is,
in its turn, based on the uniqueness theorem verifying that a
solution of an electrostatic problem satisfying the boundary
condition is the only possible solution \cite{Cheng:book}. Electric-,
and magnetic field of an infinitesimal dipole in the vicinity of an
infinite PEC surface is one of the subjects that can be studied and
facilitated by applying the method of images.
In the following section, the method of images is applied to derive the
electromagnetic modeling for different electrical sources above a
PEC surface.
\subsection{The Electric Field for Sources above a PEC Surface}\thispagestyle{empty}
It is assumed that an electric point charge $q$ is located at a
vertical distance $y=r$ above an appropriate large conducting plane
that is grounded. It will be difficult to apply the ordinary field
solution in this case but by the image methods, where an equivalent
system is presented, it will be considerably easier to solve the
original problem. An equivalent problem can be to place an image
point charge $-q$ on the opposite side of the PEC plane, i.e.
$y=-r$. In the equivalent problem, the boundary condition is not
changed and a solution to the equivalent problem will be the only
correct solution. The potential at the arbitrary point $P(x,y,z)$ is
\cite{Cheng:book2}
\begin{eqnarray}\label{eq:point_potential}
\Phi(x,y,z)=\frac{q}{4\pi\epsilon_
{0}}\left(\frac{1}{\sqrt{x^{2}+(y-r)^{2}+z^{2}}}-\frac{1}{\sqrt{x^{2}+(y+r)^{2}+z^{2}}}\right),
\end{eqnarray}
which is a contribution from both charges $q$ and $-q$ as
\begin{eqnarray}\label{eq:pos_point_potential}
\Phi_{+}(x,y,z)=\frac{q}{4\pi\epsilon_
{0}}\left(\frac{1}{\sqrt{x^{2}+(y-r)^{2}+z^{2}}}\right)
\end{eqnarray}
and
\begin{eqnarray}\label{eq:neg_point_potential}
\Phi_{-}(x,y,z)=\frac{-q}{4\pi\epsilon_
{0}}\left(\frac{1}{\sqrt{x^{2}+(y+r)^{2}+z^{2}}}\right),
\end{eqnarray}
respectively. According to the image methods, Eqn.
(\ref{eq:point_potential}) gives the potential due to an electric
point source above the PEC plane on the region $y>0$. The field
located at $y<0$ will be zero; it is indeed the region where the
image charge $-q$ is located.

Now it is assumed that a long line
charge of constant charge $\lambda$ per unit length is located at
distance $d$ from the surface of the grounded conductor, occupying half
of the entire space. It is also assumed that the line charge is
parallel to both the grounded plane and to the $z$-axis in the
rectangular coordinate system. Further, the surface of the
conducting grounded plane is coincided with $yz$-plane and $x$-axis
passes through the line charge so that the boundary condition for
this system is $\Phi(0,y,z)=0$ where $\Phi$ is defined as the
electric potential. To find the potential everywhere for this system
applying the method of images, one may start by converting this
system to an equivalent system where the boundary condition of the
original problem will be preserved. To solve this problem by the method
of images, the original system will first be converted to another
system where the conducting grounded plane vanishes, i.e. a
system where the line charge is in the free-space. By using the polar
coordinate system, the potential at an arbitrary point $P$, is
%see Fig. \ref{fig:lambda}
%, is
\begin{eqnarray}\label{eq:line_potential}
\Phi(R,\phi)=\frac{\lambda}{2\pi\epsilon_
{0}}\ln\left[\frac{({4L_{2}L_{1})^{1/2}}}{R}\right].
\end{eqnarray}
\begin{figure}[!ht]
   \centering
    \includegraphics[width=60mm]{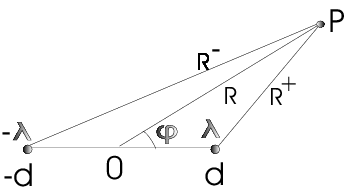}\\
    (a)\\
\end{figure}
\begin{figure}[!ht]
    \centering
    \includegraphics[width=60mm]{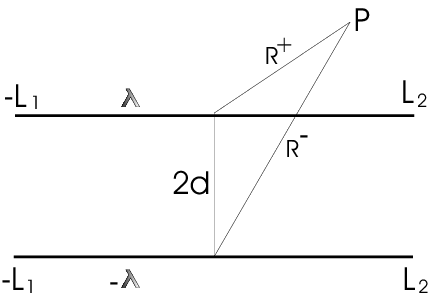}\\
    (b)\\
    \caption{Geometry of two opposite long line charges, $\lambda$ and $-\lambda$
    at distance $2d$ from each other and observed as (a):
    perpendicular to the paper plane,
    (b): coincided by the paper plane.} \label{fig:lambda}
\end{figure}
An equivalent problem may consist of a system of two parallel long
lines with opposite charges in the free-space at distance $2d$
from each other; the charge densities of the two lines are assumed
to be $\lambda$ and $-\lambda$, respectively. According to the method
of images, the total potential $\Phi$ will be determined by
contribution from these two line charges, which respectively are
\begin{figure}[!ht]
    \centering
    \includegraphics[width=60mm]{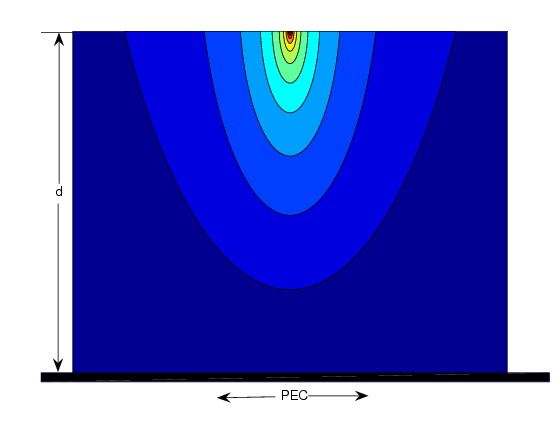}\\
    \caption{Electric potential of an infinitely long line charge parallel to the PEC surface
     at height $d$ above it.} \label{fig:phi_x}
\end{figure}
\begin{eqnarray}\label{eq:line_potential2}
\Phi^{+}=\frac{\lambda}{2\pi\epsilon_
{0}}\ln\left[\frac{({4L_{2}L_{1})^{1/2}}}{R^{+}}\right]
\end{eqnarray}
and
\begin{eqnarray}\label{eq:line_potential3}
\Phi^{-}=-\frac{\lambda}{2\pi\epsilon_
{0}}\ln\left[\frac{({4L_{2}L_{1})^{1/2}}}{R^{-}}\right].
\end{eqnarray}
The total potential is resulted from both of these two line charges
as
\begin{eqnarray}\label{eq:line_potential_4}
\Phi &=& \Phi^{+}+\Phi^{-} \nonumber\\
 &= &\frac{\lambda}{2\pi\epsilon_
{0}}\ln\left(\frac{R^{-}}{R^{+}}\right)\nonumber\\
& = & \frac{\lambda}{2\pi\epsilon_
{0}}\ln\left(\frac{d^{2}+R^{2}+2dR\cos\phi}{d^{2}+R^{2}-2dR\cos\phi}\right).
\end{eqnarray}
According to the uniqueness theorem and the method of images, Eqn.
(\ref{eq:line_potential_4}) gives the solution for a long line
charge at distance $d$ above the PEC plane. The potential below
the PEC surface will be zero. This is illustrated in Fig. \ref{fig:phi_x}.%\clearpage
\subsubsection{Radiated Electric Field of an Infinitesimal Dipole above a PEC Surface}\thispagestyle{empty}
The overall radiation properties of a radiating system can
significantly alter in the vicinity of an obstacle. The ground as a
lossy medium, i.e. $\sigma\neq0$, is expected to act as a very good
conductor above a certain frequency. Hence, by applying the method
of images the ground should be assumed as a perfect electric
conductor, flat, and infinite in extent for facilitating the
analysis. It will also be assumed that any energy from the radiating
element towards the ground undergoes reflection and the ultimate
energy amount is a summation of the reflected and directed
(incident) components where the reflected component can be accounted
for by the introduction of the image sources. In all of the
following cases, the far-field observation is considered.  To find
the electric field, radiated by a current element along the
infinitesimal length $l'$, it will be convenient to use the magnetic
vector potential $\textbf{A}$ as \cite{book:Balanis}
\begin{eqnarray}\label{eq:line_vec_pot}
\textbf{A}(x,y,z)=\frac{\mu}{4\pi}\int_{C}\textbf{I}(x',y',z')\frac{e^{-j\beta
R}}{R}dl'
\end{eqnarray}
where $(x,y,z)$ and $(x',y',z')$ represent the observation point
coordinates and the coordinates of the constant electric current
source $\textbf{I}$, respectively. $R$ is the distance from any
point on the source to the observation point; the integral path $C$
is the length of the source, and $\beta^{2}=\omega^{2}\mu\epsilon$
where $\mu$ and $\epsilon$ are permeability and permittivity of the
medium. By the assumption that an infinitesimal dipole is placed
along the $z$-axis of a rectangular coordinate system plus that it
is placed in the origin, one may write $\textbf{I}=\hat{z}I_{0}$ for
constant electric current $I_{0}$, and $x'=y'=z'=0$. Hence, the
distance $R$ will be
\begin{eqnarray}\label{eq:line_vec_pot_t}
R=\sqrt{(x-x')^{2}+(y-y')^{2}+(z-z')^{2}}=\sqrt{x^{2}+y^{2}+z^{2}}.
\end{eqnarray}
By knowing that $dl'=dz'$, and by setting
$r=\sqrt{x^{2}+y^{2}+z^{2}}$, Eqn. (\ref{eq:line_vec_pot}) may be
written as
\begin{eqnarray}\label{eq:line_vec_pot2}
\textbf{A}(x,y,z)=\hat{z}\frac{\mu I_{0}}{4\pi r}e^{-j\beta
r}\int_{-l/2}^{l/2}dz'=\hat{z}\frac{\mu I_{0}l}{4\pi r}e^{-j\beta r}.
\end{eqnarray}
The most appropriate coordinate system for studying such cases is
the spherical coordinate system, so the vector potential in Eqn.
(\ref{eq:line_vec_pot2}) should be converted into the spherical
components as
\begin{eqnarray}\label{eq:spher_vec_pot1}
A_{r}=A_{z}\cos\theta=\frac{\mu I_{0}l}{4\pi r}e^{-j\beta
r}\cos\theta,
\end{eqnarray}
\begin{eqnarray}\label{eq:spher_vec_pot22}
 A_{\theta}=-A_{z}\sin\theta=-\frac{\mu I_{0}l}{4\pi
r}e^{-j\beta r}\sin\theta,
\end{eqnarray}
\begin{eqnarray}\label{eq:spher_vec_pot3}
A_{\phi}=0.
\end{eqnarray}
In the last three equations, $A_{x}=A_{y}=0$ by the assumption that
the infinitesimal dipole is placed along the $z$-axis. For
determining the electric field radiation of the dipole, one should
operate the magnetic vector potential $\textbf{A}$ by a curl
operation to obtain the magnetic field intensity $\textbf{H}_{A}$ as
\begin{eqnarray}\label{eq:magnetic_curl1}
\textbf{H}_{A}=\frac{1}{\mu}\nabla\times\textbf{A}.
\end{eqnarray}
In spherical coordinate system, Eqn. (\ref{eq:magnetic_curl1}) is
expressed as
\begin{eqnarray}\label{eq:magnetic_curl2}
\textbf{H}_{A}=\frac{1}{\mu}\left(\hat{r}\frac{1}{r\sin\theta}
\left[\frac{\partial}{\partial\theta}(A_{\phi}\sin\theta)-
\frac{\partial A_{\theta}}{\partial A_{\phi}}\right]
+\frac{\hat{\theta}}{r}\left[\frac{1}{\sin \theta}\frac{\partial
A_{r}}{\partial\phi}-\frac{\partial}{\partial
r}(rA_{\phi})\right]+\frac{\hat{\phi}}{r}\left[\frac{\partial}{\partial
r}(rA_{\theta})-\frac{\partial
A_{r}}{\partial\theta}\right]\right).\nonumber
\end{eqnarray}
But according to Eqn. (\ref{eq:spher_vec_pot3}) and due to spherical
symmetry of the problem, where there are no $\phi$-variations along the
$z$-axis, the last equation simplifies to \cite{book:Balanis}
\begin{eqnarray}\label{eq:magnetic1}
\textbf{H}_{A}=\frac{1}{\mu}\frac{\hat{\phi}}{r}\left[\frac{\partial}{\partial
r}(rA_{\theta})-\frac{\partial A_{r}}{\partial\theta}\right],
\end{eqnarray}
which together with Eqn. (\ref{eq:spher_vec_pot1}) and
(\ref{eq:spher_vec_pot22}) gives
\begin{eqnarray}\label{eq:magnetic2}
\textbf{H}_{A}=\hat\phi\frac{\beta I_{0}l\sin\theta}{4\pi
r}j\left(1+\frac{1}{j\beta r}\right)e^{-j\beta r}.
\end{eqnarray}
Further, by equating Maxwell's equations, it will be obtained that
\begin{eqnarray}\label{eq:magnetic3}
\nabla\times\textbf{H}_{A}=\textbf{J}+j\omega\epsilon\textbf{E}_{A}.
\end{eqnarray}
By setting $\textbf{J}=0$ in Eqn. (\ref{eq:magnetic3}), it will be
obtained that
\begin{eqnarray}\label{eq:electric4}
\textbf{E}_{A}=\frac{1}{j\omega\epsilon}\nabla\times\textbf{H}_{A}.
\end{eqnarray}
Eqn. (\ref{eq:electric4}), together with Eqns.
(\ref{eq:spher_vec_pot1})-(\ref{eq:spher_vec_pot3}) yields
\begin{eqnarray}\label{eq:electric11}
E_{r}=\eta\frac{I_{0}l\cos\theta}{2\pi r^{2}}\left[1+\frac{1}{j\beta
r}\right]e^{-j\beta r},
\end{eqnarray}
\begin{eqnarray}\label{eq:electric22}
E_{\theta}=j\eta\frac{\beta I_{0}l\sin\theta}{4\pi
r}\left[1+\frac{1}{j \beta r}-\frac{1}{\beta r^{2}}\right]e^{-j\beta
r},
\end{eqnarray}
\begin{eqnarray}\label{eq:electric33}
E_{\phi}=0,
\end{eqnarray}\thispagestyle{empty}
where $\eta=E_{\theta}/H_{\phi}$ is called the intrinsic impedance
($=377\simeq120\pi$ ohms for the free-space). Stipulating for the far-field
region, i.e. the region where $\beta r >> 1$, the electric field components
$E_{\theta}$ and $E_{r}$ in Eqns.
(\ref{eq:electric11})-(\ref{eq:electric33}) can be approximated by
\begin{eqnarray}\label{eq:electric44}
E_{\theta}\simeq j\eta\frac{\beta I_{0}l\sin\theta}{4\pi
r}e^{-j\beta r},
\end{eqnarray}
\begin{eqnarray}\label{eq:electric55}
E_{r}\simeq E_{\phi}=0,
\end{eqnarray}
which is the electric far-field solution for an infinitesimal dipole
along the $z$-axis and in the spherical coordinate system. The same
procedure may be used to solve the electric field for an
infinitesimal dipole along the $x$-axis where the magnetic vector
potential $\textbf{A}$ is defined as
\begin{eqnarray}\label{eq:vector_potential_x}
\textbf{A}= \hat{\textbf{x}} \frac{\mu I_{0}le^{-j\beta r}}{4\pi r}.
\end{eqnarray}
In the spherical coordinate system, the above equation is expressed
as
\begin{eqnarray}\label{eq:vector_potential_r}
A_{r}= A_{x}\sin\theta\cos\phi,
\end{eqnarray}
\begin{eqnarray}\label{eq:vector_potential_theta}
A_{\theta}= A_{x}\cos\theta\cos\phi,
\end{eqnarray}
\begin{eqnarray}\label{eq:vector_potential_phi}
A_{\phi}= -A_{x}\sin\phi.
\end{eqnarray}
It should be mentioned that $A_y=A_z=0$ due to the placement of the
infinitesimal dipole along the $x$-axis. By far-field approximation, and
based on Eqns.
(\ref{eq:vector_potential_r})-(\ref{eq:vector_potential_phi}), the
electric field can be written as
\begin{eqnarray}\label{eq:E_r_x}
E_{r}\simeq 0,
\end{eqnarray}
\begin{eqnarray}\label{eq:E_theta_x}
E_{\theta}\simeq -j\omega A_\theta=-j\omega\frac{\mu
I_{0}le^{-j\beta r}}{4\pi r}\cos\theta\cos\phi,
\end{eqnarray}
\begin{eqnarray}\label{eq:E_phi_x}
E_{\phi}\simeq -j\omega A_\phi=-j\omega\frac{\mu I_{0}le^{-j\beta
r}}{4\pi r}\sin\phi.
\end{eqnarray}
The electric field, as a whole, will be contributions from both
$A_\theta$ and $A_\phi$ which is expressed as
\begin{eqnarray}\label{eq:E_A_x}
E_{A}\simeq -j\omega\left(A_\theta+A_\phi\right)= -j\omega\frac{\mu
I_{0}le^{-j\beta r}}{4\pi r}\left(\cos\theta\cos\phi-\sin\phi\right).
\end{eqnarray}
\subsubsection{Infinitesimal Vertical Dipole Above a PEC Surface}\thispagestyle{empty}
The overall radiation properties of a radiating system can
significantly alter in the vicinity of an obstacle. The ground as a
medium is expected to act as a very good conductor above a certain
frequency. Applying the method of images and for simplifying the
analysis, the ground is assumed to be a perfect electric conductor,
flat, and infinite in extent. It is also assumed that energy from
the radiating element undergoes reflection and the ultimate energy
amount is a summation of the reflected and the direct components
respectively where the reflected component can be accounted for by
the image sources.

A vertical dipole of infinitesimal length $l$ and
constant current $I_0$, is now assumed to be placed along $z$-axis
at distance $d$ above the PEC surface by an infinite extent. The
far-zone directed-, and reflected components in a far-field point
$P$ are respectively given by \cite{book:balanis_2}
\begin{eqnarray}\label{eq:E_d_theta}
E_{\theta}^{D}\simeq j\eta\frac{\beta I_{0}le^{-j\beta r_1}}{4\pi
r_1}\sin\theta_1,
\end{eqnarray}
and
\begin{eqnarray}\label{eq:E_r_theta}
E_{\theta}^{R}\simeq j\eta\frac{\beta I_{0}le^{-j\beta r_2}}{4\pi
r_2}\sin\theta_2,
\end{eqnarray}
where $r_1$ and $r_2$ are the distances between the observation
point and the two other points, the source- and the image-
locations; $\theta_1$ and $\theta_2$ are the related angles between
these lines and the $z$-axis. It is intended to express all the
quantities only by the elevation plane angle $\theta$ and the radial
distance $r$ between the observation point and the origin of the
spherical coordinate system. For this purpose, one may utilize the
law of cosines and also a pair of simplifications regarding the
far-field approximation. The law of cosines gives
\begin{eqnarray}\label{eq:cos_1}
r_1= \sqrt{r^{2}+d^{2}-2rd\cos\theta},
\end{eqnarray}
\begin{eqnarray}\label{eq:cos_2}
r_2= \sqrt{r^{2}+d^{2}-2rd\cos(\pi -\theta)}.
\end{eqnarray}
By binomial expansion and regarding phase variations, one may write
\begin{eqnarray}\label{eq:phase_variat_1}
r_1= r-d\cos\theta,
\end{eqnarray}
\begin{eqnarray}\label{eq:phase_variat_2}
r_2= r+d\cos\theta.
\end{eqnarray}
By utilizing the far-zone approximation where $r_1\simeq r_2\simeq
r$, and all of the above simplifications, it is obtained that
\begin{eqnarray}\label{eq:E_theta_d_r}
E_\theta^{total}=E_{\theta}^{D}+E_{\theta}^{R}= j\eta\frac{\beta
I_{0}le^{-j\beta r}}{4\pi r}\sin\theta\left(e^{+j\beta
d\cos\theta}+e^{-j\beta d\cos\theta}\right).
\end{eqnarray}
Finally, after some algebraic manipulations, one may find for $z\geq
0$
\begin{eqnarray}\label{eq:E_theta_final}
E_\theta^{total}=j\eta\frac{\beta I_{0}le^{-j\beta r}}{4\pi
r}\sin\theta\left[2\cos(\beta d\cos\theta)\right].
\end{eqnarray}
According to the image theory, the field will be zero for $z<0$.
\clearpage
\section{Computational Electromagnetics}
Determining of Green's functions for stratified media has, during
the last decades, been an important and fundamental stage to design
of high-frequency circuits. In the case of a layered medium, a
so-called \emph{mixed-potential integral equation (MPIE)}, is
applied to the associated geometry \cite{paper:Mosig}. MPIE can be
solved in both spectral-, and spatial domain and the both solutions
require appropriate Green's functions. The Green's functions for
multi-layered planar media are represented by the Sommerfeld's
integral whose integrand is consisted of the Hankel function, and
the closed-form spectral-domain Green's functions
\cite{book:W_C_Chew}. A two-dimensional inverse Fourier
transformation is needed to determine the spectral-domain Green's
functions analytically via the following integral which is along the
Sommerfeld's integration path (SIP) and the $k_{\rho}$-plane as
\begin{eqnarray}
G=\frac{1}{4\pi}\int_{SIP}dk_{\rho}k_{\rho}H_{0}^{(2)}(k_{\rho}\rho)\tilde{G}(k_{\rho})
\end{eqnarray}
where $H_{0}^{(2)}$ is the Hankel function of the second kind; $G$
and $\tilde{G}$ are the Green's functions in the spatial- and
spectral- domain. One of the topics in this context is that there is
no general analytic solution to the Hankel transform of the
closed-form spectral-domain Green's function. Numerical solution of
the above transformation integral is very time-consuming, partly due
to the slow-decaying Green's function in the spectral domain, partly
due to the oscillatory nature of the Hankel function. Dealing with
such problem constitutes one of the major topics within the
computational electromagnetics for multi-layered media. In many
applications, the \emph{Discrete complex image methods} (DCIM) is
used to handle this numerically time-consuming process. The strategy
in this process is to obtain Green's functions in a closed-form as
\begin{eqnarray}
G\cong \sum_{k=1}^{N}a_{n}\frac{e^{-jkr_{m}}}{r_{m}}
\end{eqnarray}
where
\begin{eqnarray}
r_{m}=\sqrt{\rho^{2}-b_{m}^{2}}
\end{eqnarray}
with $j=\sqrt{-1}$ will be complex-valued. The constants $a_{n}$ and
$b_{m}$ are to be determined by numerical processes such as the
Prony's method \cite{paper:Yang-Chow-Fang}\cite{book:Hildebrand}. %\clearpage
In dyadic form and by assuming an $e^{j\omega t}$ time dependence,
the electric field at an observation point, defined by the vector $\vec{r}$, produced by a
surface current $\vec{J}$ of a surface $S'$ can be expressed as
\begin{eqnarray}
\mathbf{E}(\mathbf{r})&=&-j\omega\int_{S'}\left[\overline{\overline{I}}+
\frac{1}{\beta^{2}\nabla\nabla}\right]\frac{\mu e^{-j\beta R}}{4\pi
R}\mathbf{J}(\mathbf{r},\mathbf{r}')d S' \nonumber\\
&=&
\int_{S'}\mathbf{G}(\mathbf{r},\mathbf{r}')\mathbf{J}(\mathbf{r},\mathbf{r})'d
S'
\end{eqnarray}
where $\beta=\omega\sqrt{\mu\epsilon}$ by $\mu$ and $\epsilon$ as
the electromagnetic characteristics for the layered medium; $R$ is
the distance from the source point to the field point.
$\overline{\overline{I}}$ is the unit dyad and
$\mathbf{G}(\mathbf{r},\mathbf{r}')$ is defined as the dyadic
Green's function.
There are different methods to construct the auxiliary Green's function in the case of boundary value problems,
which are as a consequence of using mathematics to study problems
arising in the real world. The numerical solution of an integral equation has the general
property that the coefficient matrix in the ultimate linear equation
$Ax=y$ will consist of a dense coefficient matrix $A$ and a
relatively fewer number of elements in the unknown vector $x$.
Numerical solution of a general integral equation involves
challenges due to the ill-conditioned coefficient matrix $A$, as a
rule and not as an exception; the integration operator to solve a differential equation is a
smoothing operator and the differential operator to solve an integral equation will be a non-smooth operator.
This is the main reason of
the ill-conditioning. Generally, and depending on the kind of
problem, there are several numerical methods to handle the
ill-conditioning and in the case of solution of Maxwell's equations
in the integral form, ill-conditioning will be a
problem to handle.\footnote{More about integral equations and ill-conditioning in the next sections.}
\subsection{Analytical Solution of Electromagnetic Fields}
Generally, the exact mathematical solution of the field problem is the most satisfactory solution, but in modern applications one cannot use such analytical solution in majority of cases. Although the analytical solution of the field problem has its limitations, the numerical methods cannot be applied without checking and realizing the limitations in classical analytical methods. Indeed, every numerical method involves an analytical simplification to the point where it is easy to apply a certain numerical method. The most commonly used analytical solutions in computational electromagnetics are

%\begin{description}
\begin{itemize}
\item Laplace, and Fourier transforms,
\item Perturbation methods,
\item Separation of variables (eigenfunction expansion method),
\item Conformal mapping,
\item Series expansion.
\end{itemize}
%\end{description}
The method of separation of variables (eigenfunction expansion method) is described in the next subsection.
\subsubsection{Eigenfunction Expansion Method}\normalsize\thispagestyle{empty}
The method of eigenfunction expansion can be applied to derive the
Green's function for partial differential equations by known
homogeneous solution. The partial differential equation
\begin{eqnarray} \label{eq:29}
U_{xx} & = & \frac{1}{\kappa }U_{t}+Q(x,t),\hspace{30mm} 0<x<L, t>0 \\
B.C. & : & U(0,t)=0, U(L,t)=0, \hspace{13mm} t>0 \nonumber \\
I.C. & : & U(x,0)=F(x), \hspace{27mm} 0<x<L \nonumber
\end{eqnarray}
with
\begin{eqnarray}
Q(x,t) & = &\frac{1}{\kappa }K_{t}(x,t)-q(x,t) \\
F(x) &=&f(x)-K(x,0)  \nonumber
\end{eqnarray}
features a problem with homogeneous boundary conditions. The Green's
function, in this case, can be represented in terms of a series of
orthonormal functions that satisfy the prescribed boundary
conditions. In this process, it is assumed that the solution of the
partial differential equation may be written in the form
\cite{book:Gunnar_Sparr}
\begin{equation} \label{eq:31}
U(x,t)=\sum\limits_{n=1}^{\infty }E_{n}(t)\Psi _{n}(x)
\end{equation}
where $\Psi_{n}(x)$ are eigenfunctions belonging to the associated
eigenvalue problem\footnote{Clearly $U(x,t)$, satisfies the
prescribed homogeneous boundary conditions, since each eigenfunction
$\Psi _{n}(x)$ does.}
\begin{equation} \label{eq:32}
X''+\lambda X=0
\end{equation}
by prescribed boundary condition (B.C.) and initial conditions
(I.C.). $E_{n}(t)$ are time-dependent coefficients to be determined.
It is also assumed that termwise differentiation is
permitted\footnote{The operation of termwise differentiation of an
infinite series is valid according to:
\underline{Corollary} If $f_{k}(x)$ has a continuous derivative on $%
[a,b] $ for each $k=1,2,3,...$ and if $\sum_{k=1}^{\infty }$
$f_{k}(x)$ converges to $S(x)$ on $[a,b]$ and if the series $\sum_{k=1}^{\infty }$ $%
f_{k}^{' }(x)$ converges uniformly to $g(x)$ on $[a,b]$ then
$S^{' }(x)=g(x)$ for every $x\in [a,b];$ equivalently $\frac{d}{dx}%
\sum_{k=1}^{\infty } $ $f_{k}(x)=\sum_{k=1}^{\infty }$ $\frac{d}{dx}%
f_{k}(x)...$''. Introduction to Mathematical Analysis page
206-William Parzynski, Philip W. Zipse.}. In this case
\begin{equation} \label{eq:33}
U_{t}(x,t)=\sum\limits_{n=1}^{\infty }E_{n}^{' }(t)\Psi _{n}(x)
\end{equation}
and
\begin{equation} \label{eq:34}
U_{xx}(x,t)=\sum\limits_{n=1}^{\infty }E_{n}(t)\Psi_{n}''(x), \nonumber
\end{equation}
which together with (\ref{eq:32}) gives
\begin{equation} \label{eq:35}
U_{xx}(x,t)=-\sum\limits_{n=1}^{\infty }\lambda _{n}E_{n}(t)\Psi
_{n}(x).
\end{equation}
This is a result of applying the superposition principle which can
be deduced as $\Psi_{n}''(x)=-\lambda _{n}\Psi
_{n}(x)$ from (\ref{eq:32}). Next, by rewriting the partial
differential equation above as
\begin{equation}
\kappa U_{xx}=U_{t}+\kappa Q(x,t)
\end{equation}
and inserting the expressions (\ref{eq:33}) and (\ref{eq:34}) into
the right-hand side of (\ref{eq:35}), it can be obtained that
\begin{equation} \label{eq:37}
\kappa U_{xx}=\sum\limits_{n=1}^{\infty }[E_{n}'(t)+\kappa
\lambda _{n}E_{n}(t)]\Psi _{n}(x).
\end{equation}
The right-hand side of the equation above is interpreted as a
generalized Fourier series\footnote{These series can be used in
developing infinite series like Fourier series and have the general
form $f(x)=\sum\limits_{n=1}^{\infty }c_{n}U _{n}(x)$ for $x_1 < x <
x_2,$ where the set of functions $\left\{ U _{n}(x)\right\} $ is
orthogonal on the specified interval by a given weighting function
$w(x)>0,$ that is \ $\int\limits_{x_{1}}^{x_{2}} U _{k}(x)U
_{n}(x)w(x)dx=0,$ \ \ for all $k\neq n.$} of the function $ \kappa
U_{xx}$ for a fixed value of $t.$ Thus, the Fourier coefficients are
defined as
\begin{eqnarray} \label{eq:38}
E_{n}^{' }(t)+\kappa \lambda _{n}E_{n}(t) &=&\kappa
\frac{1}{\left\|
\Psi _{n}(x)\right\| ^{2}}\int\limits_{0}^{L}Q(x,t)\Psi _{n}(x)dx \\
\textnormal{ for } \ n &=&1,2,...  \nonumber
\end{eqnarray}
where $\left\| \Psi _{n}(x)\right\| $ is defined as the norm of
$\Psi _{n}(x) $ with the relation
\begin{equation} \label{eq:39}
\left\| \Psi _{n}(x)\right\| ^{2}=\int\limits_{0}^{L}[\Psi _{n}(x)]^{2}dx%
\textnormal{, for } n=1,2,...
\end{equation}
Eqn. (\ref{eq:37}) as a first-order linear differential equation, has
the general solution
\begin{equation} \label{eq:40}
E_{n}(t)=\left( c_{n}+\frac{1}{\kappa }\int\limits_{0}^{t}exp(\frac{1}{%
\kappa }\lambda _{n})P_{n}(\tau )d\tau \right) exp(-\frac{1}{\kappa
}\lambda _{n}t)
\end{equation}
for $n=1,2,3,...$ by the assumption that $\lambda _{n}\neq 0$ for
all $n.$ It has to be added that $c_{n}$ are arbitrary constants. In
the equation above, $P_{n}(t)$ is defined as
\begin{equation} \label{eq:41}
P_{n}(t)=\frac{1}{\left\| \Psi _{n}(x)\right\| ^{2}}\int%
\limits_{0}^{L}Q(x,t)\Psi _{n}(x)dx \textnormal{, for } n=1,2,3,...
\end{equation}
Now, by substituting (\ref{eq:40}) into (\ref{eq:31}), it will be
obtained that
\begin{equation}
U(x,t)=\sum\limits_{n=1}^{\infty }\left( c_{n}+\frac{1}{\kappa }%
\int\limits_{0}^{t}exp(\frac{1}{\kappa }\lambda _{n})P_{n}(\tau
)d\tau \right) exp(-\frac{1}{\kappa }\lambda _{n}t)\Psi _{n}(x)
\end{equation}
For determining the arbitrary coefficients $c_{n}$, $n=1,2,3,...$,
one shall force Eqn. (\ref{eq:41}) to satisfy the prescribed
initial condition. By using the above process and applying the
method of moments (MoM), described in the previous sections, the
scattering problem of a dielectric half-cylinder which is
illuminated by a transmission wave can be obtained by the matrix
equation \cite{book:Sadiku}
\begin{eqnarray}
[A][E]=[E^i]
\end{eqnarray}
where
\begin{eqnarray}
 E^i= e^{jk(x_m\cos\phi_i+y_m\sin\phi_i)}
\end{eqnarray}
and
\begin{eqnarray}
 A_{mn}&=& \epsilon_m+j\frac{\pi}{2}(\epsilon_m-1)ka_nH_1^{(2)}(ka_m)
\hspace{7mm} for\hspace{7mm} m=n\nonumber\\
 &=& j\frac{\pi}{2}(\epsilon_m-1)ka_nJ_1^{(2)}(ka_n)H_0^{(2)}(k\rho_{mn})
 \hspace{7mm} for\hspace{7mm} m\neq n
\end{eqnarray}
with
\begin{eqnarray}
 \rho_{mn}= \sqrt{(x_m-x_n)^2+(y_m-y_n)^2}
\end{eqnarray}
for $m,n=1,2,...,N$ by $N$ as the number of cells the cylinder is
divided into. $\epsilon_m$ is the average dielectric constant of
cell $m$ and $a_m$ is the radius of the equivalent circular cell by
the same cross section as cell $m$. $E$ is the field inside the
dielectric half-cylinder and $J_1^{(2)}$ is the Bessel function
\cite{Arfken:book}; $H_1^{(2)}$ and $H_0^{(2)}$ are Hankel functions
of the first and second kinds.
\subsection{Numerical Solution of Electromagnetic Fields}
Almost any problem involving derivatives, integrals, or non-linearities cannot be solved in a finite number of steps and thus must be solved by a theoretically infinite number of iterations for converging to an ultimate solution; this is not possible for practical purposes where problems will be solved by a finite number of iterations until the answer is approximately correct. Indeed, the major aspect is, by this approach, finding rapidly convergent iterative algorithms in which the error and accuracy of the solution will also be computed. In computational electromagnetics, a difficult problem like a partial differential equation or an integral equation will be replaced by, for instance, a much simpler linear equation system. Replacing complicated functions with simple ones, non-linear problems with linear problems, high-order systems by low-order systems and infinite-dimensional spaces with finite-dimensional spaces are applied as other alternatives to solve easier problems that have the same solution to a difficult mathematical model.
%\subsubsection{Electromagnetic Modeling Techniques}
Numerical modeling of electromagnetic (EM) properties are used in, for example, the electronic industry to:
\begin{enumerate}
\item \emph{Ensure functionality of electric systems}. System performance can be degraded due to unwanted EM interference coupling into sensitive parts.
\item \emph{Ensure compliance with electromagnetic compatibility (EMC) regulations and directives}. To prevent re-designs of products and ensure compliance with directives post-production.
\end{enumerate}
The technique for solving field problems, Maxwell's equations, can be classified as experimental, analytical (exact), or numerical (approximate). The experimental techniques are expensive and time-consuming but are still used. The analytical solution of Maxwell's equations involves, among others, \emph{separation of variables} and \emph{series expansion}, but are not applicable in the general case. The numerical solution of the field problems became possible with the availability of high performance computers. The most popular numerical techniques are (1) \emph{Finite difference methods (FDM)}, (2) \emph{Finite element methods (FEM)}, (3) \emph{Moment methods (MoM)}, (4) \emph{Partial element equivalent circuit (PEEC) method}. The differences in the numerical techniques have their origin in the basic mathematical approach and therefore make one technique more suitable for a specific \emph{class of problems} compared to the others. Typical classes of problems in the area of EM modeling are:
\begin{itemize}
\item Printed circuit board (PCB) simulations (mixed circuit and EM problem).
\item Electromagnetic field strength and pattern characterization.
\item Antenna design.
\end{itemize}
Further, the problems presented above require different kinds of analysis in terms of:
 \begin{itemize}
\item Requested solution domain (time and/or frequency).
\item Requested solution variables (currents and/or voltages or electric and/or magnetic fields).
\end{itemize}
The categorization of EM problems into classes and requested solutions in combination with the complexity of Maxwell's equations emphasizes the importance of using the right numerical technique for the right problem to enable a solution in terms of accuracy and computational effort.
In the following sections, four different types of EM computational techniques are briefly presented. The first three, FEM, MoM, and FDM are the most comon techniques used today for simulating EM problems. The fourth technique, the PEEC method, is widely used within signal integrity.
\subsubsection{Finite Element Method}
 The finite element method (FEM) \cite{book:J_Jin} is a powerful numerical technique to handle problems involving complex geometries and heterogeneous media. The method is more complicated than FDM but also applicable to the wider range of problems. FEM is based on the differential formulation of Maxwell's equations in which the complete field space is discretized. The method is applicable in both the time,- and frequency domain. In this method, partial differential equations (PDEs) are solved by a transformation to matrix equations \cite{book:W_Cheney}. This is done by minimizing the energy using the mathematical concept of a functional $F$, where the energy can be obtained by integrating the (unknown) fields over the structure volume \cite{book:C_Johnson}. The procedure is commonly explained by considering the PDE described by the function $u$ with corresponding excitation function $f$ as \cite{book:B_Archambeault}\cite{book:J_Carlsson}:
 \begin{eqnarray}
 Lu = f
\end{eqnarray}
where $L$ is a PDE operator. For example, Laplace equation is given by $L = \nabla^{2}$, $u = V$, and $f = 0$. The next step is to discretize the solution region into finite elements for which the functional can be written. The functional for each FEM element, $F_{e}$, is then calculated by expanding the unknown fields as a sum of known basis functions, $u_{e_{i}}$, with unknown coefficients, $\alpha_{i}$. The total functional is solely dependent on the unknown coefficients $\alpha_{i}$ and can be written as
  \begin{eqnarray}
 F = \sum_{\forall{e}}F_{e}
\end{eqnarray}
where $e$ is the number of finite elements in the discretized structure and
\begin{eqnarray}
 F_{e} = \sum_{\forall{e}}\alpha_{i}u_{e_{i}}
\end{eqnarray}
where $i$ depends on what kind of finite elements are used in the discretization. The last step is to minimize the functional for the entire region and solve for the unknown coefficients, $\alpha_{i}$, to be zero, i.e.
\begin{eqnarray}
 {\partial F\over \partial \alpha_{i}} = 0,  \forall{i}
\end{eqnarray}
The method offers great flexibility to model complicated geometries with the use of nonuniform elements. As for the FDM, the FEM delivers the result in field variables, $\vec{E}$ and $\vec{H}$, for general EM problems at all locations in the discretized domain and at every time or frequency point. To obtain structured currents and voltages post-processing is needed for the conversion.
\subsubsection{Finite Difference Methods}
In this section a finite difference time domain (FDTD) method is described. The method is widely used within EM modeling mainly due to its simplicity. The FDTD method can be used to model arbitrarily heterogeneous structures like PCBs and the human body.
In the FDTD method finite difference equations are used to solve Maxwell's equations for a restricted computational domain. The method requires the whole computational domain to be divided, or discretized, into volume elements (cells) for which Maxwell's equations have to be solved. The volume element sizes are determined by considering two main factors \cite{book:B_Archambeault}:
\begin{enumerate}
  \item \emph{Frequency}. The cell size should not exceed $\frac{\lambda}{10}$, where $\lambda$ is the wavelength corresponding to the highest frequency in the excitation.
  \item \emph{Structure}. The cell sizes must allow the discretization of thin structures.
\end{enumerate}
The volume elements are not restricted to cubical cells, parallelepiped cells can also be used with a side to side ratio not exceeding $1:3$, mainly to avoid numerical problems \cite{book:J_Carlsson}. In many cases, the resulted FDTD method is based according to the well-known Yee formulation \cite{paper:K_S_Yee}. However, there are other FDTD methods which are not based in the Yee cell and thus have another definition of the field components. To be able to apply Maxwell's equations in differential form to the Yee cell, the time and spatial derivatives using finite difference expressions will result in the FDTD equations \cite{book:D_M_Sullivan}. The equations are then solved by:
\begin{enumerate}
  \item Calculating the electric field components for the complete structure.
  \item Advancing time by $\frac{\Delta t}{2}$.
  \item Calculating the magnetic field components for the complete structure based on the electric field components calculated in $1$.
  \item Advancing time by $\frac{\Delta t}{2}$ and continuing to $1$.
\end{enumerate}
The FDTD method delivers the result in field variables, $\vec{E}$ and $\vec{H}$, at all locations in the discretized domain and at every time point. To obtain structured currents and voltages post-processing is needed for the conversion.
\subsubsection{Method of Moments}
Method of moments (MoM) is based on the integral formulation of the Maxwell's equations \cite{book:R_F_Harrington}. The basic feature makes it possible to exclude the air around the objects in the discretization. The method is usually employed in the frequency domain but can also be applied to the time domain problems.
In the MoM, integral-based equations, describing the current distribution on a wire or a surface, are transformed into matrix equations easily solved using matrix inversion. When using the MoM for surfaces, a wire-grid approximation of the surface can be utilized as described in \cite{book:B_Archambeault}. The wire formulation of the problem simplifies the calculations and is often used for field calculations.
The starting point for theoretical derivation is to apply a linear (integral) operator, $L$, involving the appropriate Green's function $G(\vec{r},\vec{r'})$, applied to an unknown function, $I$, by an equation as \cite{book:B_Archambeault} \cite{book:R_F_Harrington}
 \begin{eqnarray} \label{eq:mom1}
 L I = f
\end{eqnarray}
where $f$ is the known excitation function for the above system. As an example the above equation can be the Pocklington's integral equation \cite{book:Sadiku}, describing the current distribution $I(z')$ on a cylindrical antenna, written as
\begin{eqnarray} \label{eq:mom2}
 \int_a^b I(z')({\partial^2 \over \partial z^2} + k^2) G(z,z')dz' = j \omega \epsilon E_{z}
\end{eqnarray}
Then the un-known function, $I$, can be expanded into a series of known functions, $u_{i}$, with un-known amplitudes, $I_{i}$, resulting in
\begin{eqnarray} \label{eq:mom3}
 I = \sum_{i=1}^n I_{i}u_{i}
\end{eqnarray}
where $u_{i}$, are called basis (or expansion) functions. To solve the unknown amplitudes, $n$, equations are derived from the combination of Eqn. (\ref{eq:mom1}) and Eqn. (\ref{eq:mom3}) and by the multiplication of $n$ weighting (or test) functions, integrating over the wire length (the cylindrical antenna) and the formulation of a proper inner product \cite{book:Sadiku}. This results in the transformation of the problem into a set of linear equations which can be written in matrix form as
\begin{eqnarray} \label{eq:mom4}
 [Z][I] = [V]
\end{eqnarray}
where the matrices, $Z$, $I$, and $V$ are referred to as generalized impedance, current, and voltage matrices and the desired solution for the current, $I$, is obtained by matrix inversion. Thus, the unknown solution is expressed as a sum of known basis functions whose weighting coefficients corresponding to the basis functions will be determined for the best fit. The same process applied to differential equations is known as the "weighted residual" method \cite{book:Peterson}. The MoM delivers the result in system current densities $\vec{J}$ and/or voltages at all locations in the discretized structure and at every frequency point (depending on the integral in Eqn. (\ref{eq:mom2})). To obtain the results in terms of field variables, post-processing is needed for the conversion.
The well-known computer program \emph{Numerical Electromagnetics Code}, often referred to as NEC \cite{homepage:NEC}, utilizes the MoM for calculation of the electromagnetic response for antennas and other metal structures.
\subsubsection{The Method of Partial Element Equivalent Circuit}
The basis of the Partial Element Equivalent Circuit (PEEC) method originates from inductance calculations performed by Dr. Albert E. Ruehli at IBM T.J. Watson Research Center, during the first part of 1970s \cite{paper:Ruehli1}\cite{paper:Ruehli2}\cite{paper:Ruehli3}. Dr. Ruehli was working with electrical interconnect problems and understood the benefits of breaking a complicated problem into basic partitions, for which inductances could be calculated to model the inductive behavior of the complete structure \cite{paper:Ruehli1}\cite{paper:Ruehli4}. By doing so, return current paths need not to be known \emph{a priori} as required for regular (loop) inductance calculations.
The concept of partial calculations was first introduced by Rosa \cite{bulletin:Rosa} in 1908, further developed by Grover \cite{paper:Grover} in 1946 and Hoer and Love \cite{paper:Hoer} in 1965. However, Dr. Ruehli included the theory of partial coefficients of potential and introduced the partial element equivalent circuit (PEEC) theory in 1972 \cite{paper:Ruehli5}. Significant contributions of the PEEC method includes:
\begin{itemize}
  \item The inclusion of dielectrics \cite{paper:Ruehli6},
  \item The equivalent circuit representation with coefficients of potential \cite{paper:Heeb1},
  \item The retarded partial element equivalent circuit representation \cite{paper:Heeb2},
  \item PEEC models to include incident fields, scattering formulation \cite{paper:Ruehli7},
  \item Nonorthogonal PEECs \cite{paper:Ruehli8}.
\end{itemize}
The interest and research effort of the PEEC method have increased during the last decade. The reasons can be an increased need for combined circuit and EM simulations and the increased performance of computers enabling large EM system simulations. This development reflects on the areas of the current PEEC research, for example, model order reduction (MOR), model complexity reduction, and general speed up.
%
%\section{Basic PEEC Theory}
%
The PEEC method is a 3D, full wave modeling method suitable for
combined electromagnetic and circuit analysis.  In the PEEC method,
the integral equation is interpreted as the Kirchhoff's voltage law
applied to a basic PEEC cell which results in a complete circuit
solution for 3D geometries.  The equivalent circuit formulation
allows for additional SPICE-type circuit elements to easily be
included.  Further, the models and the analysis apply to both the
time and the frequency domain. The circuit equations resulting from
the PEEC model are easily constructed using a condensed modified
loop analysis (MLA) or modified nodal analysis (MNA) formulation
\cite{Al:MNA75}. In the MNA formulation, the volume cell currents
and the node potentials are solved simultaneously for the
discretized structure. To obtain field variables, post-processing
of circuit variables are necessary. This section gives an outline
of the nonorthogonal PEEC method as fully detailed in
\cite{paper:Ruehli8}.  In this formulation, the objects, conductors and
dielectrics, can be both orthogonal and non-orthogonal quadrilateral
(surface) and hexahedral (volume) elements. The formulation
utilizes a global and a local coordinate system where the global
coordinate system uses orthogonal coordinates $x,y,z$ where the global
vector $\vec{F}$ is of the form $ \vec{F} = {F_x} \vec{\hat{x}} +
{F_y} \vec{\hat{y}} + {F_z} \vec{\hat{z}}$.  A vector in the global
coordinates are marked as $\vec{r_g}$.  The local coordinates
$a,b,c$ are used to separately represent each specific possibly
non-orthogonal object and the unit vectors are $\vec{\hat{a}}$,
$\vec{\hat{b}}$, and $\vec{\hat{c}}$, see further \cite{paper:Ruehli8}.
The starting point for the theoretical derivation is the total
electric field on the conductor expressed as
\begin{equation} \label{eq:Efield}
    \vec{E}^i (\vec{r_g},t) =
   \frac{\vec{J}(\vec{r_g},t)}{\sigma} + \frac {\partial
   \vec{A}(\vec{r_g},t)}{\partial t} + \nabla \phi (\vec{r_g},t),
\end{equation}
where $\vec{E}^i$ is the incident electric field, $\vec{J}$ is the
current density in a conductor, $\vec{A}$ is the magnetic vector
potential, $\phi$ is the scalar electric potential, and $\sigma$ is the
electrical conductivity.  The dielectric areas are taken into
account as an excess current with the scalar potential using the
volumetric equivalence theorem.  By using the definitions of the
vector potential $\vec{A}$ and the scalar potential $\phi$ we can
formulate the integral equation for the electric field at a point
$\vec{r_g}$ which is to be located either inside a conductor or
inside a dielectric region according to
\begin{eqnarray}
   \label{eq:Fullie}
   {\vec{E}^{i}}(\vec{r_g},t)
    & \hspace{-2mm} = & \hspace{-2mm}
   \frac{\vec{J}(\vec{r_g},t)}{\sigma}  \\
    & \hspace{-2mm} + & \hspace{-2mm}
   {\mu} \int_{v'} G(\vec{r_g},\vec{r_g}') \frac {\partial
\vec{J}(\vec{r_g}',{t_d})}{\partial t} {d v'} \nonumber \\ %
    & \hspace{-2mm} + & \hspace{-2mm}
    \epsilon_0 (\epsilon_r \hspace{-0.5mm} - \hspace{-0.5mm} 1)
   {\hspace{0.5mm}\mu} \hspace{-1mm} \int_{v'} \hspace{-1mm} G
(\vec{r_g}, \vec{r_g'})
   \frac {\partial^2 \vec{E}(\vec{r_g}', t_d)}{\partial {t^2}}
\nonumber \\ %
   & \hspace{-2mm} + & \hspace{-2mm}
    \frac {\nabla}{\epsilon_0}  \int_{v'} G
(\vec{r_g},\vec{r_g}')  {q}(\vec{r_g}',{t_d})  d v'
.
\nonumber %
\end{eqnarray}
Eqn. (\ref{eq:Fullie}) is the time domain formulation which can
easily be converted to the frequency domain by using the Laplace
transform operator $s = \frac{\partial}{\partial t}$ and where the
time retardation $\tau$ will transform to $e^{-s \tau}$.
\begin{figure}
    \centering
    \includegraphics[width=0.5\linewidth]{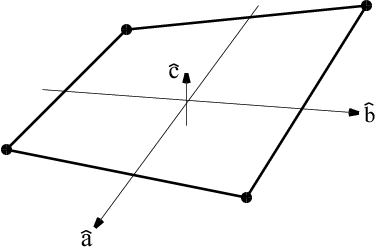}\\
    \caption{Nonorthogonal element created by the mesh generator with
associated local coordinate system.}\label{fig:Basic_c}
\end{figure}
The PEEC integral equation solution of the Maxwell's equations is based
on the total electric field, e.g. (\ref{eq:Efield}).  An integral or
inner product is used to reformulate each term of~(\ref{eq:Fullie})
into the circuit equations.  This inner product integration converts
each term into the fundamental form $\int \vec{E} \cdot dl = V$
where $V$ is the voltage or potential difference across the circuit
element. It can be shown how this transforms the sum of the electric
fields in~(\ref{eq:Efield}) into the Kirchhoff's Voltage Law (KVL) over
a basic PEEC cell. Fig. \ref{fig:PEECmo_o} details
the ($L_p$,$P$,$\tau$)PEEC model for the metal patch in Fig.
\ref{fig:Basic_c} when discretized using four edge nodes (solid dark
circles). The model in Fig. \ref{fig:PEECmo_o} consists of: %
\begin{itemize}
\item partial inductances ($L_p$) which are calculated from the
volume cell discretization using a double volume integral.
\item coefficients of potentials ($P$) which are calculated from the
surface cell discretization using a double surface integral.
\item retarded controlled current sources, to account for
the electric field couplings, given by $I_p^i =
\frac{p_{ij}}{p_{ii}} I_C^j (t-t_{d_{ij}})$ where $t_{d_{ij}}$ is
the free space travel time (delay time) between surface cells $i$
and $j$,
\item retarded current controlled voltage sources, to account for
the magnetic field couplings, given by $V_L^n = Lp_{nm}
\frac{\partial \, I_m (t-t_{d_{nm}})}{\partial t},$ where
$t_{d_{nm}}$ is the free space travel time (delay time) between
volume cells $n$ and $m$.
\end{itemize}
\begin{figure*}
    \centering
    \includegraphics[width=0.6\linewidth]{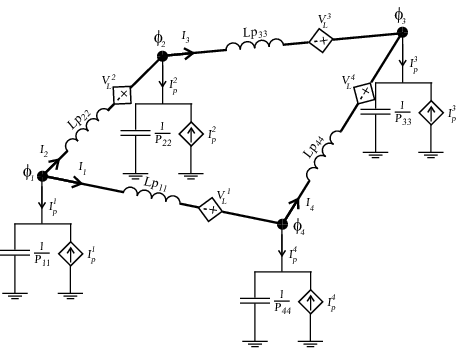}\\
    \caption{($L_p$,$P$,$\tau$)PEEC model for metal patch in
Fig.~\ref{fig:Basic_c}
    discretized with four edge nodes. Controlled current sources, $I_p^n$,
account
    for the electric field coupling and controlled voltage sources, $V_L^n$,
account
    for the magnetic field coupling.}
    \label{fig:PEECmo_o}
\end{figure*}
By using the MNA method, the PEEC model circuit elements can be
placed in the MNA system matrix during evaluation by the use of
correct matrix stamps \cite{Al:MNA75}. The MNA system, when used to
solve frequency domain PEEC models, can be schematically described
as
\begin{equation}  \label{eq:MNA_A}
\begin{array}{ccc}
j \omega \emph{\textbf{P}}^{-1}V -\emph{\textbf{A}}^T I & = & I_s\\%
\emph{\textbf{A}}V - (\emph{\textbf{R}} + j \omega
\emph{\textbf{L}}_p) I &
= & V_s %
\end{array}
\end{equation}
where: \emph{\textbf{P}} is the coefficient of potential matrix,
\emph{\textbf{A}} is a sparse matrix containing the connectivity
information, \emph{\textbf{L}$_p$} is a dense matrix containing the
partial inductances, elements of the type $Lp_{ij}$,
\emph{\textbf{R}} is a matrix containing the volume cell
resistances, \emph{{V}} is a vector containing the node potentials
(solution), elements of the type $\phi_{i}$, \emph{{I}}  is a vector
containing the branch currents (solution), elements of the type
$I_{i}$, \emph{{I}$_s$}  is a vector containing the current source
excitation, and \emph{{V}$_s$} is a vector containing the voltage
source excitation. The first row in the equation system in
(\ref{eq:MNA_A}) is the Kirchhoff's current law for each node while the
second row satisfy the Kirchhoff's voltage law for each basic PEEC cell
(loop). The use of the MNA method when solving PEEC models is the
preferred approach since additional active and passive circuit
elements can be added by the use of the corresponding MNA stamp. For
a complete derivation of the quasi-static and full-wave PEEC circuit
equations using the MNA method, see for example \cite{Garr:PhD}.
%\begin{thebibliography}{1}
%\%bibitem{Al:NonoJ03} A. E. Ruehli \emph{et al}., ``Nonorthogonal PEEC
%formulation for time- and frequency-domain modeling". \emph{IEEE
%Trans. on EMC}, 45(2):167-176, May 2003.
%\bibitem{Garr:PhD} J. E. Garrett, ``Advancements of the Partial Element
%Equivalent Circuit Formulation", PhD dissertation, The University of
%Kentucky, 1997.
%\bibitem{Al:MNA75} C. Ho, A. Ruehli and P. Brennan, ``The modified nodal
%approach to network analysis", \emph{IEEE Trans. on Circuits and
%Systems}, pages 504--509, June 1975.
%\end{thebibliography}
%\makebib
%\end{bibunit}
%%%%%%%%%%%%%%%%%%%%%%%%%%%%%%%%%%%%%%%%%%%%%%%%%%%%%%%
%\subsubsection{PEEC Current Density Expansion}
%%%%%%%%%%%%%%%%%%%%%%%%%%%%%%%%%%%%%%%%%%%%%%%%%%%%%%%
\clearpage
\section{Medical Diagnostics and Microwave Tomographic imaging by Applying Electromagnetic Scattering}
The main objective of this section is to investigate biological imaging algorithms by solving the direct, and inverse electromagnetic scattering problem due to a model based illustration technique within the microwave range. A well-suited algorithm will make it possible to fast parallel processing of the heavy and large numerical calculation of the inverse formulation of the problem. The parallelism of the calculations can then be performed and implemented on GPU:s, CPU:s, and FPGA:s. By the aid of mathematical/analytical methods and thereby faster numerical algorithms, an improvement of the existing algorithms is also expected to be developed. These algorithms may be in time domain, frequency domain and a combination of both.

There is a potential in the microwave tomographic imaging for providing information about both physiological state and anatomical structure of the human body. By several strong reasons the microwave tomographic imaging is assumed to be tractable in medical diagnostics: the energy in the microwave region is small enough to avoid ionization effects in comparison to X-ray tomography. Furthermore, tissue characteristics such as blood content, blood oxygenation, and blood temperature cannot be differentiated by the density-based X-ray tomography. The microwave tomography can be used instead of determining tissue properties by means of complex dielectric values of tissues. It is shown that the microwave tissue dielectric properties are strongly dependent on physiological condition of the tissue \cite{article:Serguei Y. Semenov et al}. The dependence of the tissue dielectric properties plays a major roll to open opportunities for microwave imaging technology within medical diagnostics. As in tomography by X-ray densities of tissues are investigated, the electromagnetic scattering technique is based on determining the permittivity of tissues. In such context, the interesting thing to think about is, always, how the old electromagnetic scattering computations can be improved by smarter faster mathematical/numerical algorithms. In addition, there are promising methods providing a good compromise between rapidity and cost why there is a potential interest of microwave imaging in biomedical applications. The area of the research is rather new so that new approaches and new methods are expected to be developed in tomographic imaging.

The inverse electromagnetic scattering should be solved in order to produce a tomographic image of a biological object. In this process, the dielectric properties of the object under test is deduced from the measured scattered field due to the object and a known incident electric field. Nonlinearity relations arise between the scattered field and multiple paths through the object. Approximations are used to linearize the resulting nonlinear inverse scattering problem. As this problem is ill-posed, the existence and uniqueness of the solution and also its stability should be established \cite{Tommy Henriksson:PhD}.
\subsection{The Direct Electromagnetic Scattering Problem}
Scattering theory has had a major roll in the twentieth century mathematical physics. The theory is concerned with the effect an inhomogeneous medium has on an incident particle or wave. The direct scattering problem is to determine a scattered field $u^{s}$ from a knowledge of an incident field $u^{i}$ and the differential equation governing the wave equation. The incident field is emitted from a source , an antenna for example, against an inhomogeneous medium. The total field is assumed to be the sum of the incident field $u^{i}$ and the scattered field $u^{s}$. The governing differential equation in such cases is Maxwell's equations which will be converted to the wave equation. Generally, the direct scattering problems depend heavily on the frequency of the wave in question. In particular, the phenomenon of diffraction is expected to occur if the wavelength $\lambda = 2 \pi/k$ is very small compared to the smallest observed distance; $k$ is the wave number. Thus, due to the scattering obstacle, an observable shadow with sharp edges is produced. Obstacles which are small compared with the wavelength disrupt the incident wave without any identifiable shadow. Two different frequency regions are therefore defined based on the wave number $k$ and a typical dimension of the scattering objects $a$. The set of $k$ values such that $ka >> 1$ is called the \emph{high frequency region} and the set of $k$ values where $ka \leq 1$ is called the \emph{resonance region}. The distinction between these two frequency regions is due to the fact that the applied mathematical methods in the resonance region differ greatly from the methods used in the high frequency region.

One of the first issues to think about when studying the direct scattering problem is the \emph{uniqueness} of the solution. Then, by having established uniqueness, the existence of the solution and a numerical approximation of the problem must be analyzed and handled. The uniqueness of the solution will be discussed in the next section.
\subsubsection{Uniqueness of the Solution}
Within the electromagnetic field theory there are two fundamental governing differential equations for electrostatics in any medium. These are \cite{Cheng:book2}:
\begin{eqnarray}
  \nabla \cdot \mathbf{D} = \rho_{v},  \label{eq:div.D}\\
   \nabla \times \mathbf{E} = 0, \label{eq:rotD}
\end{eqnarray}
where $\mathbf{D}$ and $\mathbf{E}$, are the electric flux density and electric field intensity, as defined earlier; $\rho_{v}$ is the volume charge density. Because $\mathbf{E}$ is rotation-free, a scalar electric potential $V$ can be defined such that
\begin{equation}\label{eq:scalar V}
   \mathbf{E} = - \nabla \Phi.
\end{equation}
Combining (\ref{eq:div.D}) and (\ref{eq:scalar V}) yields
\begin{equation}\label{eq:scalar2}
   \nabla\cdot(\epsilon\nabla \Phi) = -\rho_{v}
\end{equation}
where $\epsilon$ is the permittivity due to linear isotropic medium in which $\mathbf{D} = \epsilon \mathbf{E}$. The above equations will finally result in
\begin{equation}\label{eq:poisson}
   \nabla^{2} \Phi = \frac{-\rho_{v}}{\epsilon}.
\end{equation}
Eqn. (\ref{eq:poisson}) is called the \emph{Poisson's equation}. In this equation $\nabla^{2}$ is \emph{Laplacian}. If there is no charge in the simple medium, i.e. $\rho_{v}= 0$, then Eqn. (\ref{eq:poisson}) will be converted into
\begin{equation}\label{eq:laplace}
   \nabla^{2} \Phi = 0,
\end{equation}
which is called the \emph{Laplace's equation}. The concept of uniqueness has arisen when solving the Laplace's,- or Poisson's equation by different methods. Depending on the complexity and the geometry of the problem, one may use analytical, numerical, or experimental methods. The question is whether all of these methods will give the same solution. This may be reformulated as: Is the present particular solution of the Laplace's,- or Poisson's equation, satisfying the boundary conditions, the only solution? The answer will be yes by relying on the uniqueness theorem. Irrespective of the method, a solution of the problem satisfying the boundary conditions is the only possible solution.

In connection with the concept of the uniqueness, two theorems are extensively discussed within the computational electromagnetics \cite{Arfken:book}. These are:
\begin{theorem} \label{theorem:unique1}
A vector is uniquely specified by giving its divergence and its curl within a simply connected region and its normal component over the boundary.
\end{theorem}
\begin{theorem} \label{theorem:unique2}
A vector $\mathbf{V}$ with both source and circulation densities vanishing at infinity may be written as the sum of two parts, one of which is irrotational, the other solenoidal.
\end{theorem}
A proof of the uniqueness theorem due to the Laplace's equation is given in \cite{book:sadiku2}. The theorem (\ref{theorem:unique2}) is called the Helmholtz's theorem. The theorems (\ref{theorem:unique1}) and (\ref{theorem:unique2}) can together be interpreted as: "\emph{a solution of the Poisson's equation (\ref{eq:poisson}) and Eqn. (\ref{eq:laplace}) (as a special case), which satisfies a given boundary condition, is a unique solution}" \cite{Cheng:book2}. In \cite{book:Balanis}, there is another interpretation of the uniqueness theorem:

"\emph{A field in a lossy region is uniquely specified by the sources within the region plus the tangential components of the electric field over the boundary, or the tangential components of the magnetic field over the boundary, or the former over part of the boundary and the latter over the rest of the boundary}".
Hence, according to the uniqueness theorem, the field at a point in space will be sufficiently determined by having information about the tangential electric field and the tangential magnetic field on the boundary. This means that to determine the field uniquely, one of the following alternatives must be specified \cite{TsangDing1:book}:
\begin{itemize}
  \item $\hat{n}\times\hat{E}$ everywhere on $S$,
  \item $\hat{n}\times\hat{H}$ everywhere on $S$,
  \item $\hat{n}\times\hat{E}$ on a part of $S$ and $\hat{n}\times\hat{H}$ on the rest of $S$,
\end{itemize}
with $S$ as the boundary of the domain. Directly related to the electromagnetic obstacle scattering two other theorems can be found in \cite{book:D.Colton}; these are:
\begin{theorem} \label{theorem:scatterObstacle1} Assume that $D_{1}$ and $D_{2}$ are two perfect conductors such that for one fixed wave number the electric far-field patterns for both scatterers coincide for all incident directions and all polarizations. Then $D_{1} = D_{2}$.
\end{theorem}
\begin{theorem} \label{theorem:scatterObstacle2} Assume that $D_{1}$ and $D_{2}$ are two perfect conductors such that for one fixed incident direction and polarization the electric far field patterns of both scatterers coincide for all wave numbers contained in some interval $0 < k_{1} < k < k_{2} < \infty$. Then $D_{1} = D_{2}$.
\end{theorem}
%
%Theorem (\ref{theorem:scatterObstacle1}) is the uniqueness result for the direct formulation whilst (\ref{theorem:scatterObstacle2}) is a theorem for fixed direction and polarization.
As depicted in the above theorems, the scattered wave depends analytically on the wave number $k$.
\subsubsection{Solution of the Direct Electromagnetic Scattering Problem}
The simplest problem in the direct scattering problem is scattering by an impenetrable obstacle D. Then, the total field $u$ can be determined by \cite{book:D.Colton}
\begin{eqnarray}
   \nabla^{2}u + k^{2}n(x)u = 0  \hspace{2mm} in \hspace{2mm}\mathbb{R}^{3}, \\
   u(x) = e^{ikx\cdot d} + u^{s}(x), \\
   \lim_{r \to \infty} r(\frac{\partial u^{s}}{\partial r} - iku^{s}) = 0, \label{eq:boundary1}
 %\nonumber \\ %
    %& \hspace{-2mm} + & \hspace{-2mm}
%\nonumber %
\end{eqnarray}
in which $r = |x|$, and $n = c_{0}^{2}/c^{2}$ is the refractive index due to the square of the sound speeds. By the assumption that the medium is absorbing and also assuming that $1 - n$ has \emph{compact support}\footnote{See Appendix A.}, $n$ will be complex-valued. For the homogeneous host medium, $c = c_{0}$, and for the inhomogeneous medium, $c = c(x)$. Depending on obstacle properties, different boundary conditions will be assumed. Eqn. (\ref{eq:boundary1}) is called \emph{Sommerfeld radiation condition}. Acoustic wave equations possessing such kind of boundary condition guarantee that the scattered wave is outgoing.

Within the computational electromagnetics for the scattering problem, the incident field by the time-harmonic electromagnetic plane wave can be expressed as
\begin{eqnarray}
   E^{i}(x,t) =  i k (d\times p)\times d e^{i(kx\cdot d -\omega t)}  \\
   H^{i}(x,t) =  i k (d\times p)e^{i(kx\cdot d -\omega t)}
   %\hspace{2mm} in \hspace{2mm}\mathbb{R}^{3} \backslash \bar{D}
 %\nonumber \\ %
    %& \hspace{-2mm} + & \hspace{-2mm}
%\nonumber %
\end{eqnarray}
where $k=\omega \sqrt{\epsilon_{0}\mu_{0}}$ is the wave number, $\omega$ the radial frequency, $\epsilon_{0}$ the electric permittivity in vacuum, $\mu_{0}$ the magnetic permeability in vacuum, $d$ the direction of propagation and $p$ the polarization. Assuming variable permittivity but constant permeability, the electromagnetic scattering problem is now to determine both the electric, and magnetic field according to
\begin{eqnarray}\label{eq:EM-scatterng_1}
   \nabla \times E - i k H = 0 \hspace{3mm} \mbox{in} \hspace{2mm} \mathbb{R}^{3} \\ \nonumber
   \nabla \times H + i k n(x)E = 0 \hspace{3mm} \mbox{in} \hspace{2mm} \mathbb{R}^{3} %\mathds{R}
\end{eqnarray}
where $n = \varepsilon / \varepsilon_{0}$ is the refractive index by the ratio of the permittivity $\epsilon = \epsilon (x)$ in the inhomogeneous medium and and $\varepsilon_{0}$ the permittivity in the homogeneous host medium; $n$ will have a complex value if the medium is conducting. It is assumed that $1 - n$ has compact support. The total electromagnetic field is determined by
\begin{eqnarray} \label{eq:EM-scattering_3}
   E(x) = (i/k)\nabla \times \nabla \times p e^{ikx\cdot d} + E^{s}(x) \\
   H(x) = \nabla \times pe^{ikx\cdot d} + H^{s}(x) \label{eq:EM-scattering_4}  %in \mathds{R},
 %\nonumber \\ %
    %& \hspace{-2mm} + & \hspace{-2mm}
%\nonumber %
\end{eqnarray}
so that
\begin{equation} \label{eq:Silver-M}
   \lim_{r\rightarrow \infty} (H^{s} \times x - r E^{s}) = 0
 %\nonumber \\ %
    %& \hspace{-2mm} + & \hspace{-2mm}
%\nonumber %
\end{equation}
where Eqn. (\ref{eq:Silver-M}) is called the \emph{Silver-M\"{u}ller radiation condition}.
%\begin{equation}
 %  u(x) = e^{ikx.d} + u^{s}(x)
 %\nonumber \\ %
    %& \hspace{-2mm} + & \hspace{-2mm}
%\nonumber %
%\end{equation}
%
The electromagnetic scattering by a perfect obstacle $D$ is now to find an electromagnetic field such that \cite{book:D.Colton}
\begin{eqnarray}\label{eq:HarmMaxw}
   \nabla E - i k H = 0, \hspace{2mm} \nabla H - i k E = 0 \hspace{3mm} \mbox{in} \hspace{2mm} \mathbb{R}^{3}\setminus\bar{D},\\
   E(x) = (i/k)\nabla\times\nabla \times p e^{ikx\cdot d} + E^{s}(x)   \\
   H(x) = \nabla\times pe^{ikx\cdot d} + H^{s}(x)  \\
   \nu\times\nabla E = 0 \hspace{3mm} \mbox{on} \hspace{2mm} \partial D, \\
   \lim_{r\rightarrow \infty} (H^{s} \times x - r E^{s}) = 0,
 %\nonumber \\ %
    %& \hspace{-2mm} + & \hspace{-2mm}
%\nonumber %
\end{eqnarray}
where $\nu$ is the unit outward normal to $\partial D$. Eqns. (\ref{eq:HarmMaxw}) are called the \emph{time harmonic Maxwell's equations}. The above formulation is called the \emph{direct electromagnetic scattering problem}. The method of integral equations is a common method to investigate the existence of a numerical approximation of the direct problem. The integral equation associated with the electromagnetic scattering problem due to Eqns.(\ref{eq:EM-scatterng_1})-(\ref{eq:EM-scattering_4}) is given by \cite{book:D.Colton}
\begin{eqnarray} \label{eq:EM-scattering_5}
   E(x) = \frac{i}{k}\nabla \times \nabla \times pe^{ikx\cdot d} - k^{2}\int_{\mathbb{R}^3} \Phi (x,y) m(y) E(y) \\ \nonumber
   + \nabla \int_{\mathbb{R}^3} \frac{1}{\nu (y)} \nabla n(y)\cdot E(y) \Phi(x,y) dy, \hspace{3mm} x \in \mathbb{R}^{3},
 %\nonumber \\ %
    %& \hspace{-2mm} + & \hspace{-2mm}
%\nonumber %
\end{eqnarray}
where
\begin{equation} \label{eq:EM-scattering_6}
   \Phi(x,y) := \frac{1}{4\pi}\frac{e^{ik\mid x-y\mid}}{\mid x-y \mid}, \hspace{2mm} x\neq y,
 %\nonumber \\ %
    %& \hspace{-2mm} + & \hspace{-2mm}
%\nonumber %
\end{equation}
and $m:=1-n$; if $E$ is the solution of Eqn. (\ref{eq:EM-scattering_6}), one can define
\begin{equation} \label{eq:EM-scattering_7}
   H(x) := \frac{1}{ik}\nabla \times E(x)
\end{equation}
Letting $x$ tend to the boundary of $D$ and introducing $a$ as a tangential density to be determined, one can verify that $a$ will be a solution for $E$ in the following boundary integral equation \cite{book:D.Colton}:
\begin{eqnarray} \label{eq:EM-Scattering_8}
  E^{s}(x) = \nabla \times \int_{\partial D} a(y) \Phi(x,y) ds(y), \hspace{2mm} x \in \mathbb{R}^{3}\backslash\bar{D}\\ \nonumber
   H^{s}(x) = \frac{1}{ik}\nabla \times E^{s}(x), \hspace{2mm} x \in \mathbb{R}^{3} \backslash \bar{D}.
\end{eqnarray}
%
%\begin{equation} \label{eq:EM-scattering_8}
 %  \Phi(x,y) := \frac{1}{4\pi}\frac{e^{ik\mid x-y\mid}}{\mid x-y \mid}, \hspace{2mm} x\neq y,
 %\nonumber \\ %
    %& \hspace{-2mm} + & \hspace{-2mm}
%\nonumber %
%\end{equation}
%
In this formulation, the boundary integral equation in Eqns. (\ref{eq:EM-Scattering_8}) will be used to solve Eqns. (\ref{eq:EM-scatterng_1})-(\ref{eq:EM-scattering_4}). The fact is that the integral equation is not uniquely solvable if $k^{2}$ is a Neumann eigenvalue of the negative Laplacian in $D$ \cite{book:K. Atkinson}. The numerical solution of boundary integral equations in scattering theory is generally a much challenging area and a deeper understanding of this topic requires knowledge in different areas of functional analysis, stochastic processes, and scientific computing. In fact, the electromagnetic inverse medium problem is not entirely investigated and numerical analysis and experiments have yet to be done for the three dimensional electromagnetic inverse medium.
 %However, a theoretical basis of the dual space method for electromagnetic waves has already been developed %\cite{article:Colton and P\ddot{a}iv\ddot{a}rinta} \cite{article:Colton and Kress} \cite{article:Colton %and H\ddot{a}hner}.
%%%%%%%%%%%%%%%%%%%%%%%%%%%%%%%%%%%%%%%%%%%%%%%%%%%%%%%%
\subsection{The Inverse Electromagnetic Scattering Problem}
The inverse scattering problem is, in many areas, of equal interest as the direct scattering problem. Inverse formulation is applied to a daily basis in many disciplines such as image and signal processing, astrophysics, acoustics, geophysics and electromagnetic scattering. The inverse formulation, as an interdisciplinary field, involves people from different fields within natural science. To find out the contents of a given black box without opening it, would be a good analogy to describe the general inverse problem. Experiments will be carried on to guess and realize the inner properties of the box. It is common to call the contents of the box "the model" and the result of the experiment "the data". The experiment itself is called "the forward modeling." As sufficient information cannot be provided by an experiment, a process of regularization will be needed. The reason to this issue is that there can be more than one model ('different black boxes') that would produce the same data. On the other hand, improperly posed numerical computations will arise in the calculation procedure. A regularization process in this context plays a major roll to solve the inverse problem.
\subsubsection{Analytical Formulation of the Inverse Scattering Problem}
%A properly concise formulation of the inverse scattering problem can be found in \cite{book:Gerhard K.} in which
As in the direct formulation, the permittivity $\epsilon$ has a constant value, in inverse scattering formulation $\epsilon$ has to be assumed as room-dependent. Assuming $\epsilon = 1$ outside a sphere with radius $R$, and $\epsilon \neq 1$ inside, the following equation can be deduced by starting from Maxwell's equations and some vector algebra \cite{book:Gerhard K.}
\begin{equation} \label{eq:scattInverse1}
  \nabla\times(\nabla \times \mathbf{E}(\mathbf{r},\omega))-\omega^2\epsilon_{0}\mu_{0}\epsilon(\mathbf{r})\mathbf{E}(\mathbf{r},\omega) = \mathbf{0}
\end{equation}
where $\mathbf{r}$ is the room variable and the scatterer material with volume $V_{s}$ is assumed to be non-magnetic, i.e. $\mu=1$; no other current sources except induced current generated by the incident field $\mathbf{E}^i$ are assumed to exist either. By introducing a dimensionless quantity $\chi_{e}$, known as the \emph{electric susceptibility}, a new equation will be introduced as
\begin{equation} \label{eq:chi}
  \mathbf{D} = \epsilon_{0}(1+\chi_{e}(\mathbf{r}))\mathbf{E}(\mathbf{r},\omega)\\
            =\epsilon_{0}\epsilon(\mathbf{r})\mathbf{E}(\mathbf{r},\omega)=\epsilon \mathbf{E}(\mathbf{r},\omega)
\end{equation}
where $\mathbf{D} (C/m^2)$ is defined as \emph{electric displacement}, see previous sections. By Eqn. (\ref{eq:chi}), it is easy to see that
\begin{equation} \label{eq:chi2}
  \epsilon(\mathbf{r}) = 1+\chi_{e}(\mathbf{r})=\frac{\epsilon}{\epsilon_{0}}.
\end{equation}
A dielectric medium is, by definition, linear if $\chi$ is independent of $\mathbf{E}$ and homogeneous if $\chi$ is independent of space coordinates. In fact, the electric susceptibility $\chi$ gives the dielectric deviation between the free-space and other dielectric media in the case of inverse scattering problem. It is equal to zero in the free-space on the outside of the sphere with radius $R$ and distinct from zero inside. The sphere contains in fact the scatterer with the volume $V_{s}$. In addition, it is assumed that the medium contained in the volume $V_{s}$ is not \emph{dispersive}, i.e. $\chi$ inside the volume $V_{s}$ is not dependent on the frequency $\omega$. In the case of the inverse electromagnetic scattering problem, the goal is to determine the function $\chi(\mathbf{r})$ by experimentally obtained incident electric field $E^i$ and scattered electric field $E^s$ and the total field $E=E^i+E^s$. This process is started by re-writing the Eqn. (\ref{eq:scattInverse1}) as
\begin{equation} \label{eq:scattInverse2}
  \nabla\times(\nabla \times \mathbf{E}(\mathbf{r},\omega))-k^2\mathbf{E}(\mathbf{r},\omega)= k^2 \chi_{e}(\mathbf{r})\mathbf{E}(\mathbf{r},\omega)
\end{equation}
where
\begin{equation} \label{eq:scattInverse_k^2}
  k^2 = \omega^2 \epsilon_{0}\mu_{0}
\end{equation}
in which $k$ is the wave number associated with vacuum as the surrounding medium. Due to the incident field $\mathbf{E}^i$, a current will be induced in $V_{s}$ with the associated current density $\mathbf{J}_s$, which can be expressed as \cite{book:Gerhard K.}
\begin{equation} \label{eq:ScatteredE^s}
  \mathbf{J}_{s} = -j\omega \epsilon_{0} \chi_{e}\mathbf{E}.
\end{equation}
By the aid of this induced current density, the scattered electric field can be expressed as \cite{book:Gerhard K.}
\begin{equation} \label{eq:IncCurrentDensity}
  \mathbf{E}^s(\mathbf{r}) = [k^2+\bigtriangledown\bigtriangledown]\cdot\int_{V_{s}}\frac{e^{jk|\mathbf{r}-\mathbf{r}'|}}{4\pi|\mathbf{r}-\mathbf{r'}|}\chi_{e}(\mathbf{r}')\mathbf{E}(\mathbf{r}')dv', \hspace{4mm} \mathbf{r} \not \in V_{s}.
\end{equation}
As it is seen in Eqn. (\ref{eq:IncCurrentDensity}), the integral deals with the inside of the scatterer which is unobservable by experimentally measuring the electric field. Both the scattered,- and the incident electric field can be measured at the outside of the scatterer and the unknown electric field inside the integral should be determined in different situations. In the cases where $\mathbf{E}^s<<\mathbf{E}^i$, there are different methods to approximate the integral in Eqn. (\ref{eq:IncCurrentDensity}). In the \emph{Born} approximation, the dielectrical properties of the scatterer can be determined by a three-dimensional inverse Fourier transforming of the far-field $\mathbf{F}$ in certain directions and for any frequency. This means that for the experimentally given incident plane wave with propagation vector $\mathbf{\hat{k}}_{i}$
\begin{equation} \label{eq:IncEField}
  \mathbf{E}^i(\mathbf{r}) =  \mathbf{E}_{0}e^{jk\hat{k}_{i}\cdot\mathbf{r}}
\end{equation}
and for a fixed point $k$, a three-dimensional Fourier transform of the function $\chi_{e}$ can be calculated in a point $k(\mathbf{\hat{k}}_{i}-\mathbf{\hat{r}})$, that is \cite{book:Gerhard K.}
\begin{equation} \label{eq:Incid_Fourier}
  \int_{V_{s}}\chi_{e}(\mathbf{r}')e^{jk(\mathbf{\hat{k}}_{i}-\mathbf{\hat{r}})\cdot \mathbf{r}'} dv'=\frac{4\pi}{k^3}\frac{\mathbf{F(\hat{\mathbf{r}})}}{\hat{\mathbf{r}}\times(\mathbf{E}_{0}\times \hat{\mathbf{r}})}
\end{equation}
where the far-field scattering amplitude (measured data in the far-field) is
\begin{equation} \label{eq:IncidFarField}
  \mathbf{F}(\mathbf{\hat{r}})=\mathbf{\hat{r}}\times(\mathbf{E}_{0}\times\mathbf{\hat{r}}) \frac{k^3}{4\pi}\int_{V_{s}}\chi_{e}(\mathbf{r}')e^{jk(\mathbf{\hat{k}}_{i}-\mathbf{\hat{r}})\cdot \mathbf{r}'} dv'.
\end{equation}
As depicted in Eqn. (\ref{eq:Incid_Fourier}), in the Born approximation the problem is linearized with substitution of the unknown field in the integral by the given incident filed. In the \emph{Rytov} approximation, the polarization field is assumed to be almost unchanged and the phase of the field is interpreted as all the scattering, that is
\begin{equation} \label{eq:Rytov1}
  \mathbf{E}(\mathbf{\hat{r}}) = \mathbf{E}_{0} e^{jk\psi(\mathbf{r})}
\end{equation}
where $\psi(\mathbf{r})$ is the field phase as
\begin{equation} \label{eq:Rytov2}
  \psi(\mathbf{r}) = \mathbf{\hat{k}}_{i}\cdot\mathbf{r} + \psi_{s} (\mathbf{r})
\end{equation}
in which $\psi_{s} (\mathbf{r})$ is the deviations from $\mathbf{\hat{k}}_{i}$, i.e., the phase associated with the incident field. By application of some vector algebra and by the aid of an approximation, (\ref{eq:scattInverse2}) can be written as \cite{book:Gerhard K.}
\begin{equation} \label{eq:Rytov3}
  2\mathbf{E}_{0}(\mathbf{\hat{k}}_{i}\cdot\nabla\psi(\mathbf{r}))-(\mathbf{E}_{0}\cdot\nabla\psi_{s}(\mathbf{r}))\mathbf{\hat{k}}_{i} = \chi_{e}(\mathbf{r})\mathbf{E}_{0}
\end{equation}
that yields
\begin{eqnarray} \label{eq:Rytov4}
  2\mathbf{\hat{k}}_{i}\cdot\nabla\psi(\mathbf{r}) = \chi_{e}(\mathbf{r}) \\ \nonumber
  \mathbf{E}_{0}\cdot\nabla\psi_{s}(\mathbf{r}) = 0
\end{eqnarray}
by which the electric susceptibility $\chi_{e}$ can be determined by the following process. \\
By introducing new Cartesian coordinates $\xi$ and $\eta$ it will be possible to have the directions of $\hat{k}_{i}$ lying in, for example, the $xy$-plane so that the $\eta\xi$-plane is perpendicular to the $xy$-plane, that is
\begin{eqnarray} \label{eq:Rytov5}
  \xi = x\cos\phi + y\sin\phi, \\ \nonumber
  \eta = -x\sin\phi + y\cos\phi,
\end{eqnarray}
where $\phi$ is the rotation angle between the two coordinate systems of $xy$ and $\eta\xi$. Finally, the phase $\psi_{s}$ can, by the Rytov approximation, be expressed as \cite{book:Gerhard K.}
\begin{equation} \label{eq:Rytov6}
  \psi_{s}(\xi,\phi) = \frac{1}{2}\int_{-\infty}^{+\infty} \chi_{e}(x,y)d\eta.
\end{equation}
There are two methods to obtain $\chi_{e}(x,y)$ from Eqn. (\ref{eq:Rytov6}): the method of \emph{Projection} and the method of \emph{Integral Equation}. Following, the method of Projection is briefly explained.

The general inverse formulation of determining dielectric properties $f(x,y)$ of the scatterer is in the form of the following integral \cite{book:Gerhard K.}
\begin{equation} \label{eq:projection1}
  u_{\phi}(\xi) = \int_{-\infty}^{\infty}f(x,y)d\eta = \int_{-\infty}^{\infty}\int_{-\infty}^{\infty}f(x,y)\delta(\xi-\mathbf{\rho}\cdot{\mathbf{\hat{\xi}}})dxdy
\end{equation}
where $\rho=\hat{\mathbf{x}}x+\hat{\mathbf{y}}y$ is a two-dimensional regional vector; Eqn. (\ref{eq:projection1}) is, by inspection, according to the definition of the Dirac's delta function $\delta$. The coordinates $\xi$ and $\eta$ are associated with the directions $\hat{\xi}$ and $\hat{\eta}$ according to
\begin{eqnarray} \label{eq:xi_eta}
  \hat{\xi} = \hat{\mathbf{x}}\cos\phi + \hat{\mathbf{y}}\sin\phi, \\ \nonumber
  \hat{\eta} = -\hat{\mathbf{x}}\sin\phi + \hat{\mathbf{y}}\cos\phi. \\ \nonumber
\end{eqnarray}
According to this formulation of inverse electromagnetic scattering, the data is actually the Fourier transform $\mathfrak{F}$ of the dielectric properties of the scatterer in question. This means
\begin{equation} \label{eq:F_transformDielctric}
  \mathfrak{F}\{{{u_{\phi}(p)}}\} = \hat{u}_{\phi}(p)=\int_{-\infty}^{\infty}u_{\phi}(\xi)e^{ip\xi}d\xi,
\end{equation}
which together with (\ref{eq:projection1}) gives
\begin{equation} \label{eq:projection2}
  \hat{u}_{\phi}(p) = \int_{-\infty}^{\infty}e^{ip\xi}\int_{-\infty}^{\infty}\int_{-\infty}^{\infty}f(x,y)\delta(\xi-\mathbf{\rho}\cdot{\mathbf{\hat{\xi}}})dxdyd\xi.
\end{equation}
By using the Dirac's delta function properties, (\ref{eq:projection2}) can be written as
\begin{equation} \label{eq:projection3}
  \hat{u}_{\phi}(p) = \int_{-\infty}^{\infty}\int_{-\infty}^{\infty}f(x,y)e^{ip\rho\cdot\hat{\xi}}dxdy.
\end{equation}
The unknown dielectric properties $f(x,y)$ can now be determined by inverse Fourier transforming of (\ref{eq:projection3}), that is \cite{book:Gerhard K.}
\begin{equation} \label{eq:projection4}
  f(x,y) = \frac{1}{4\pi^2}\int_{-\infty}^{\infty}\int_{-\infty}^{\infty}\hat{f}(\mathbf{p})e^{-j\rho\cdot\mathbf{p}}dp_{x}dp_{y}
\end{equation}
where
\begin{equation} \label{eq:projection5}
  \hat{f}(\mathbf{p}) = \hat{u}_{\phi}(p), \hspace{4mm} \mathrm{for} \hspace{2mm} p\geq 0.
\end{equation}
Expressed in the Cartesian coordinates, the vector $\mathbf{p}$ can be written as
\begin{equation} \label{eq:projection6}
 \mathbf{p} = \mathbf{\hat{x}}p_{x} + \mathbf{\hat{y}}p_{y}
\end{equation}
\subsubsection{Numerical Solution of the Inverse Electromagnetic Scattering Problem}
As the direct scattering problem has been thoroughly investigated, the inverse scattering problem has not yet a rigorous mathematical/numerical basis. Because the nonlinearity nature of the inverse scattering problem, one will face improperly posed numerical computation in the inverse calculation process. This means that, in many applications, small perturbations in the measured data cause large errors in the reconstruction of the scatterer. Some regularization methods must be used to remedy the ill-conditioning due to the resulting matrix equations. Concerning the existence of a solution to the inverse electromagnetic scattering one has to think about finding approximate solutions after making the inverse problem stabilized. A number of methods is given to solve the inverse electromagnetic scattering problem in which the nonlinear and ill-posed nature of the problem are acknowledged. Earlier attempts to stabilize the inverse problem was via reducing the problem into a linear integral equation of the first kind. However, general techniques were introduced to treat the inverse problems without applying an integral equation.

The process of regularization is used at the moment when selection of the most reasonable model is on focus. Computational methods and techniques ought to be as flexible as possible from case to case. A computational technique utilized for small problems may fail totally when it is used for large numerical domains within the inverse formulation. New methodologies and algorithms would be created for new problems since existing methods are insufficient. This is the major character of the existing inverse formulation in problems with huge numerical domains. There are both old and new computational tools and techniques for solving linear and nonlinear inverse problems. Linear algebra has been extensively used within linear and nonlinear inverse theory to estimate noise and efficient inverting of large and full matrices. As different methods may fail, new algorithms must be developed to carry out nonlinear inverse problems. Sometimes, a regularization procedure may be developed for differentiating between correlated errors and non-correlated errors. The former errors come from linearization and the latter from the measurement. To deal with the nonlinearity, a local regularization will be developed as the global regularization will deal with the measurement errors. There are researchers who have been using integral equations to reformulate the inverse obstacle problem as a nonlinear optimization problem.

In some approaches, a priori is assumed such that enough information is known about the unknown scattering obstacle $D$ \cite{Article:KirschKress1}\cite{Article:KirschKress2}\cite{Article:KirschKress3}. Then, a surface $\Gamma$ is placed inside $D$ such that $k^{2}$ is not a Dirichlet eigenvalue of the negative Laplacian for the interior of $\Gamma$. Then, assuming a fixed wave number $k$ and a fixed incident direction $d$, and also by representing the scattered field $u^{s}$ as a single layer potential \cite{book:D.Colton}
\begin{equation} \label{eq:inverse1}
  u^{s}(x) = \int_{\Gamma} \phi(y) \Phi(x,y)ds(y)
\end{equation}
where $\phi \in L^{2}(\Gamma)$ is to be determined; $L^{2}(\Gamma)$ is the space of all \emph{square integrable functions}\footnote{More about square integrable functions in Appendix A.} on the boundary $\Gamma$. The far field pattern $u_{\infty}$ is then represented as
\begin{equation} \label{eq:inverse2}
  u_{\infty}(\hat{x};d) = \frac{1}{4 \pi}\int_{\Gamma} e^{-ik\hat{x}\cdot y}\phi(y) \Phi(x,y)ds(y), \hspace{2mm} \hat{x} \in \Omega
\end{equation}
where $\Omega$ is the unit sphere, and $\hat{x}= x/|x|$. By the aid of the given (measured) far field pattern $u_{\infty}$, one can find the density $\phi$ by solving the ill-posed integral equation of the first kind in Eqn. (\ref{eq:inverse2}). This method is described thoroughly in \cite{Article:KressZinn1}\cite{Article:KressZinn2}\cite{Article:KressZinn3}.

In another method it is assumed that the given (measured) far field $u_{\infty}$ for all $\hat{x}$, and $d \in \Omega$ is given. The problem is now to determine a function $g\in L^{2}(\Omega)$ such that
\begin{equation} \label{eq:inverse3}
  \int_{\Omega}u_{\infty}(\hat{x};d) g(d) ds(d) = \frac{1}{k i^{p+1}} Y_{p}(\hat{x}), \hspace{2mm} \hat{x} \in \Omega
\end{equation}
where $p$ is an integer and $k$ as fixed; $Y_{p}$ is a spherical harmonic of order $p$ \cite{Arfken:book}. It can be shown that solving the ill-posed integral equation (\ref{eq:inverse3}) leads, in special conditions, to the nonlinear equation \cite{book:D.Colton}
\begin{equation} \label{eq:inverse4}
  \int_{\Omega}e^{ikr(a)a\cdot d} g(d) ds(d) = -h_{p}^{(1)}(kr(a))Y_{p}(\hat{x}), \hspace{2mm} a \in \Omega
\end{equation}
in which $r$ is to be determined, and where $x(a)=r(a)a$; $h_{p}^{1}$ is the spherical Hankel function of the first kind of order $p$ \cite{Arfken:book}. In \cite{Article:Blohbaum}, this method is developed and applied to the case of the electromagnetic inverse obstacle problem.
%%%%%%%%%%%%%%%%%%
%The method of Generalized Cross Validation (GCV) is recommended for noise estimation and Krylov space methods are used to invert large systems.
\subsection{Optimization of the Inverse Problem}
A linear inverse problem can be given in form of finding $\textbf{x}$ such that $\textbf{A}x= \textbf{b} + \textbf{n}$, where $\textbf{b}$, $\textbf{x}$, and $\textbf{n}$ are vectors, and $A$ is a matrix; $\textbf{n}$ is the noise which has to be minimized by different so called \emph{regularization} methods. Within the field of image processing, a forward model is defined as an unobservable input $x^*$ which returns as an observable output $\textbf{b}$. Here, the forward problem is modeled by a forward model and the inverse problem will be an approximation of $x^*$ by $\hat{x}$. The forward process is, in other words, a mapping from the image to error-free data, $\bar{d}$, and the actual corrupted data, $d$; the noise $n$ is the difference $\bar{d} - d$. The corruption in such context is due to small round off error by a computer representation and also by inherent errors in the measurement process.

The collection of values that are to be reconstructed is referred to as the \emph{image}. Denoting $f$ as the image, the forward problem is the mapping from the image to the quantities that can be measured. By the forward mapping denoted by $A$, the actual data $d$ can be denoted by
\begin{equation} \label{eq:forward}
  d = A(f) + n
\end{equation}
in which $A$ may be either a linear,- or a nonlinear mapping. Accordingly, the inverse problem can now be interpreted as finding the original image given the data, and the information from the forward problem.
\subsubsection{Well-posed and Ill-posed Problems}
As the image and data are infinite-dimensional (continuous) or finite-dimensional (discrete), there will be several classifications. Image and data can be both continuous; they can also be both discrete, or the former continuous, the latter discrete, and vice versa. However, each of the cases is approximated by a discrete-discrete alternative as computer implementation is in a discrete way. The other mentioned alternatives are always an idealization of the problem. According to Hadamard \cite{book:Ghosh and Couchman}, the inverse problem to solve
\begin{equation} \label{eq:hadamard}
A(f) = d
\end{equation}
is a \emph{well-posed} problem if
\begin{itemize}
  \item a solution exists for any data $d$,
  \item there is a unique solution in the image space,
  \item the inverse mapping from $d$ to $f$ is continuous.
\end{itemize}
In addition, an \emph{ill-posed} problem is where an inverse does not exist because the data is outside the range of $A$. Other interpretations of the above three conditions is \emph{an ill-posed problem is a problem in which small changes in data will cause large changes in the image}.
To stabilize the solution of ill-conditioned and rank-deficient problems, the concept of \emph{singular value decomposition (SVD)} is widely used. The reason is that relatively small singular values can be dropped which makes the process of computation less sensitive to perturbations in data. Another important application of the SVD is the calculation of the condition number of a matrix which is directly related to ill-posed problems.
\subsubsection{Singular Value Decomposition}
In connection with rank-deficient and ill-posed problems, it is convenient to describe singular value expansion of a kernel due to an integral equation. This calculation is by means of the singular value decomposition (SVD). All the difficulties due to ill-conditioning of a matrix will be revealed by applying SVD.
Assuming $A\in\mathbb{R}^{m\times n}$ be a rectangular or square matrix and letting $m \succeq n$, the SVD of $A$ is a decomposition in form of
\begin{equation} \label{eq:svd1}
  A = U\Sigma V^{T} = \sum_{i=1}^n u_{i}\sigma_{i}v_{i}^T
\end{equation}
where the orthonormal matrices $U = (u_{1},...,u_{n})\in\mathbb{R}^{m\times n}$ and $V = (v_{1},...,v_{n}\in\mathbb{R}^{n\times n})$ are such that $U^TU = V^TV = I_{n}$ \cite{book:Heath}. The diagonal matrix $\Sigma = diag(\sigma_{1},...,\sigma_{n})$ has decreasing nonnegative elements such that
\begin{equation} \label{eq:svd22}
  \sigma_{1}\geq \sigma_{2} \geq ... \geq \sigma_{n}\geq 0.
\end{equation}
where the vectors $u_{i}$ and $v_{i}$ are the \emph{left and right singular vectors} of $A$, respectively; $\sigma_{i}$ are called the \emph{singular values} of $A$ which are, in fact, the nonnegative square roots of the eigenvalues of $A^T A$. Columns of $U$ and $A$ are orthonormal eigenvectors of $AA^T$ and $A^T A$ respectively. The rank of a matrix is equal to the number of nonzero singular values, and a singular value of zero indicates that the matrix in question is rank-deficient. One of the most significant applications of matrix decomposition by SVD is within parallel matrix computations. The SVD has other important applications within the area of scientific computing. Some of them are as follows \cite{book:Heath}:
\begin{itemize}
  \item solving linear least squares of ill-conditioned and rank-deficient problems,
  \item calculation of orthonormal bases for range and null spaces,
  \item calculation of condition number of a matrix,
  \item calculation of the Euclidean norm.
\end{itemize}
As an example, the Euclidean norm of a matrix can be calculated by SVD as the first element in (\ref{eq:svd22}), i.e. $\sigma_{1}$. This value is indeed the first (and the largest) singular value, positioned on the diagonal matrix $\Sigma$, that is:
\begin{equation} \label{eq:2norm}
  \sigma_{max} = \|A\|_{2} = \max_{x\neq 0} \frac{\|Ax\|_{2}}{\|x\|_{2}}.
\end{equation}
With respect to the Euclidean norm in (\ref{eq:2norm}), and also the smallest singular value, both calculated by the SVD procedure, one can determine the condition number of the matrix $A$ by
\begin{equation} \label{eq:condNumb}
  cond(A) = \frac{\sigma_{max}}{\sigma_{min}}
\end{equation}
with $\sigma_{min}$ as the smallest element on the diagonal matrix $\Sigma$ in (\ref{eq:svd1}).
\subsection{Regularization}
With an origin in the \emph{Fredholm integral equation} of the first kind as \cite{Arfken:book}
\begin{equation} \label{eq:fredholm1}
  f(x) = \int_{a}^b K(x,t) \phi(t)dt
\end{equation}
with $f(x)$ and $K(x,t)$ known and $\phi(t)$ unknown, most inverse problems describe the continuous world. The \emph{kernel} $K$ represents the response functions of an instrument (determined by known signals), and $f$ represents measured data; $\phi$ represents the underlying signal to be determined. Integral equations can also result from \emph{the method of Green's functions} \cite{book:Roach} and the \emph{boundary element methods} \cite{book:Kythe} for solving differential equations. The \emph{existence} and \emph{uniqueness} of solutions to integral equations is more complicated in comparison to algebraic equations. In addition, the solution may be highly sensitive to perturbations in the input data $f$. The reason to sensitivity lies in the nature of the problem that has to do with determining the integrand from the integral; this is just the opposite integration operator which is a smoothing process. Such an integral operator with a smooth kernel $K$, i.e. a kernel that does not possess singularities, has zero as an eigenvalue \cite{book:Heath}. This means that there are nonzero functions that will be annihilated under the integral operator. Solving for $\phi$ in (\ref{eq:fredholm1}) tends to introduce high-frequency oscillation as the integrand contains $\phi$ as an arbitrary function and the smooth kernel $K$. The sensitivity in the process of solving integral equations of type (\ref{eq:fredholm1}) is inherent in the problem and it has not to do with the method of solving. For an integral operator with a smooth kernel by having zero as an eigenvalue, additional information may be required. The reason to this is that using a more accurate quadrature rule leads to a resulting ill-conditioned linear equation system, which thereby results into a more erratic solution.
To handle the ill-conditioning in such context, several numerical methods have been used. In \emph{truncated singular value decomposition} the solution of the ultimate linear equation system $Ax = y$ is computed by using the singular value decomposition of $A$. In this process, small singular values of $A$ are omitted from the solution; the small singular values of $A$ reflects and generates in fact ill-conditioning when solving the ultimate linear equation system.

The method of \emph{regularization} solves a minimization problem to obtain a physically meaningful solution. Starting from the Fredholm integral equation in (\ref{eq:fredholm1}) and introducing $m(t)$ as the model and letting $b = [b_{1},...,b_{n}]^T$ be the vector of the measured data, a connection between $m$ and $b$ will be
\begin{equation} \label{eq:fredholm2}
  b_{i} = \int_{D} K(s_{i},t) m(t)dt + \epsilon_{i}
\end{equation}
where $K(s,t)$ is still the smooth kernel, and $\epsilon_{i}$ is the measurement noise; $D$ is the domain of the integration. The goal is now to find the model $m$ assuming that the noisy data $b$ is given. The problem (\ref{eq:fredholm2}) becomes a well-posed least-squares system if it will be discretized with a number of parameters $M$ which is smaller than $N$. As a disadvantage, this discretization makes the solution lie in a small subspace which does not always fit the problem. However, by choosing a discretization with a number of parameters $M$ bigger than $N$, the discrete system will possess some of the characteristics of the continuous system.

Two different methods have been used to discretize Eqn. (\ref{eq:fredholm2}) \cite{PhD:Wolfgang Stefan}. The first method uses a quadrature rule to approximate the integral in Eqn. (\ref{eq:fredholm2}), that is
\begin{equation} \label{eq:discrFred21}
  \int_{D} K(s_{j},t) m(t)dt \approx \sum_{i=1}^M w_{i}K(s_{j},t)m(t_{i})\triangle (t_{i}).
\end{equation}
This discretization results into a rectangular system like
\begin{equation} \label{eq:discrFred22}
  b = Ax + \epsilon
\end{equation}
in which $A_{ji} = w_{i}K(s_{j},t_{j})$ and $x = m(t_{i})$ which is a vector in $\mathbb{R}^M$. The second method uses discretization by the \emph{Galerkin} methods in which the model $m$ is described by
\begin{equation} \label{eq:model_m}
  m = \sum_{i=1}^M x_{i}\psi_{i}(s)
\end{equation}
where $\psi_{i}(s)$ for $i = 1,2,...,m$ is an orthonormal set of basis functions, see Appendix. The integral in Eqn. (\ref{eq:discrFred21}) can now be written as
\begin{equation} \label{eq:discrFred2Galerk1}
  \int_{D} K(s_{j},t) m(t)dt = \sum_{i=1}^M x_{i}\int_{D} K(s_{j},t) \psi_{i}(t)dt,
\end{equation}
which is in the same form as in Eqn. (\ref{eq:discrFred22}), that is $b = Ax + \epsilon$, in which $x$ is a vector of coefficients and
\begin{equation} \label{eq:discrFred2Galerk2}
  A_{ji} =\int_{D} K(s_{j},t) \psi_{i}(t)dt.
\end{equation}
The "trade-off" is of importance to think about when selecting discretization methods in computational work; as quadrature methods are easier to implement, the Galerkin method gives more accurate results and requires fewer unknowns to obtain the same accuracy. However, the major issue to think about in this stage is that the matrix $A$ is, as a rule, ill-conditioned and to get rid of ill-conditioning, regularization is needed for the solution of the problem. In the following section, two different methods for regularization are presented. They are the \emph{Tikhonov} regularization and regularization by the \emph{subspace} methods.
\subsubsection{Tikhonov Regularization}
According to Tikhonov, the problem of finding $x$ as a solution to $b = Ax + \epsilon$ can be substituted by a minimization problem as \cite{PhD:Wolfgang Stefan}
\begin{equation} \label{eq:tikhonMin}
  \mathrm{min} \hspace{4mm} \phi(\beta,x) = \|Ax-b\|^2 + \beta\|Wx\|^2
\end{equation}
where $\phi(\beta,x)$ is called the \emph{global objective function}. In this formulation $\|Ax-b\|^2$ is the \emph{data misfit} and $\|Wx\|^2$ is called the \emph{model objective function}; $\beta$ is a \emph{penalty parameter} as a parameter that determines how well the solution is fitted with data. By adjusting $\beta$, the solution will fit the data in an optimal way. By differentiating the problem in (\ref{eq:discrFred2Galerk2}) with respect to $x$ and setting the differentiation to zero, a solution will be achieved, that is
\begin{equation} \label{eq:tikhDiff}
  (A^T A + \beta W^T W)x = A^T b.
\end{equation}
%
%It is shown that the system in (\ref{eq:tikhDiff}) is equivalent to
It is shown that the penalty parameter $\beta$ is found by solving
\begin{equation} \label{eq:penaltyBeta}
   \|b-Ax(\beta)\|^2 = \|(I-A(A^T A + \beta I)^{-1} A^T)b\|^2
\end{equation}
where $I$ is the identity matrix. Inversion or decomposition of the term $(A^T A + \beta I)^{-1}$ is costly in this equation and this constitutes a major challenge in finding the solution. In the context of inverse problems, the \emph{Tikhonov} regularization is used to damp the singular vectors, which are associated with small singular values in the problem, formulated as a singular value decomposition \cite{PhD:Wolfgang Stefan}. Referred to Eqn. (\ref{eq:penaltyBeta}) and with the matrix $A$ decomposed by singular value decomposition as
\begin{equation} \label{eq:svd2}
   A = U\Sigma V^T
\end{equation}
one can find out that
\begin{equation} \label{eq:svd3}
   (A^T A + \beta I)x = (V\Sigma^2 V^T + \beta I)x = V(\Sigma^2 + \beta I)V^T x = V\Sigma U^T b.
\end{equation}
By multiplying both sides in $V^T$ in (\ref{eq:svd3}) and by other simplifications, $x$ can be found as
\begin{equation} \label{eq:svd_x1}
   x = V\Sigma^{-1}(\Sigma^2 + \beta I)^{-1}\Sigma^2 U^T b.
\end{equation}
By having (\ref{eq:svd_x1}) in vector form, it can be written as
\begin{equation} \label{eq:svd_x2}
   x = \sum_{i=1}^N \frac{\lambda_{i}^2}{\lambda_{i}^2+\beta}\frac{b^T u_{i}}{\lambda_{i}}v_{i}.
\end{equation}
By introducing a function $f_{T}(\lambda)$  as
\begin{equation} \label{eq:filterFunction}
   f_{T}(\lambda) =  \frac{\lambda^2}{\lambda^2+\beta}
\end{equation}
which is called the \emph{Tikhonov filter function}, Eqn. (\ref{eq:svd_x2}) will be rewritten as
\begin{equation} \label{eq:svd_x3}
   x = \sum_{i=1}^N f_{T}(\lambda_{i})\frac{b^T u_{i}}{\lambda_{i}}v_{i}.
\end{equation}
In fact, the Tikhonov filter function in (\ref{eq:filterFunction}), "filters" the singular vectors which are associated with small singular values \cite{PhD:Wolfgang Stefan}. These vectors are in their turn associated with $\lambda^2$ which are much smaller than $\beta$ as the penalty parameter. The Tikhonov regularization is a fundamental process in inverse problems.
%After finding the penalty parameter $\beta$,
%Eqn. (\ref{eq:penaltyBeta}) is closely related to \emph{constraint minimization} so that Eqn. (\ref{eq:penaltyBeta}) can be formulated as
%
%\begin{eqnarray} \label{eq:tikhonConstr}
 % \mathrm{minimize} \hspace{4mm} \|Wx\|^2 \\ \nonumber
%\\ \nonumber
 %\mathrm{subject\hspace{2mm} to} \hspace{4mm} \|Ax-b\|^2 \leq T \\  \nonumber
%\end{eqnarray}
%
%where $T$ is defined as a target misfit.
\subsubsection{Subspace Regularization}
For more efficiency, the Tikhonov regularization can be extended by the \emph{Subspace} regularization method. In fact, the Tikhonov regularization solutions require a long time and considerable memory. Any shortcut like discretizing the problem with fewer parameters, leads to an overdetermined system for a solution to $b=Ax+\epsilon$. As a consequence, a coarse discretization will not fit the problem as the solution is forced into a small subspace \cite{PhD:Wolfgang Stefan}. The challenge in such context will be to transform the problem into a small appropriate one by choosing a new subspace $S_{k}$ in the minimization problem of
\begin{eqnarray} \label{eq:minimization2}
   \mathrm{minimize}\hspace{2mm}\|Ax-b\|^2  \\ \nonumber
\mathrm{Subject}\hspace{2mm}\mathrm{to} \hspace{2mm} x\in S_{k}
\end{eqnarray}
where $A:R^M\longrightarrow R^N$. Subspace regularization is involved with definition of the $k-$dimensional subspace $S_{k}$ for $k<N$ such that $S_{k}= Span(V_{k})$. Hence, the original problem of (\ref{eq:minimization2}) is now converted into an equivalent minimization problem of the least-square system of
\begin{equation} \label{eq:minimization3}
   AV_{k}z=0.
\end{equation}
In fact, a more realistic formulation in this context is to solve a minimization problem of (\ref{eq:minimization2}) by defining a subspace $S_{k}$ with $k<<N<M$ that leads to a well-posed overdetermined system by choosing a small enough $k$ and a good choice of $S_{k}$. There are different methods in which the subspace is chosen such that it is spanned by singular values.
%\section{Biological Imaging Applying Electromagnetic Scattering}
\subsection{Experimental Applications in Biological Imaging}
A two-dimensional prototype microwave tomographic imaging system composed of $64$ antennas (a circular antenna array) with the operating frequency in $2450$ MHz is considered in \cite{article:Serguei Y. Semenov et al}. The antennas are located on the perimeter of a cylindrical microwave chamber with an internal diameter of $360$ mm which can be filled with various solutions, including deionized water. By separating the antennas into emitters and receivers, the influence of the emitter signal is assumed to be avoided. The sequential radiation emitting of $32$ emitters, and $16-20$ receiving antennas, is measured. The antennas are used with a narrow radiation pattern in the vertical direction for creating a two-dimensional slice of the three-dimensional object under test (OUT). Special waveguides are also used to get a wider horizontal projection. The amplitude and the phase of the scattered field due to the OUT is also measured.

For the two-dimensional mathematical formulation it is assumed that the OUT with the complex dielectric permittivity $\epsilon = \epsilon' + \epsilon''$ is not dependent on the $z$ coordinate in the media. The OUT is located in the media with a constant complex dielectric permittivity of $\epsilon_{0} = \epsilon_{0}' + \epsilon_{0}''$. In addition, the magnetic permeability is assumed to be constant everywhere. The dielectric properties of the OUT which is assumed to be an infinite cylindrically symmetric object with volume $V$ is investigated. The situation is finally modeled by the following integral equation \cite{article:Serguei Y. Semenov et al}
\begin{eqnarray} \label{eq:OUT1}
 \frac{j}{k^2 - k_{0}^2} - \int_{v} G j d V = E^i, \hspace{3mm} \mathrm{inside} \hspace{1mm} \mathrm{V} \\ \nonumber
 \int_{v} G j d V = E^s, \hspace{3mm}  \mathrm{outside} \hspace{1mm}\mathrm{V} \nonumber
\end{eqnarray}
where
\begin{eqnarray} \label{eq:OUT2} \nonumber
 k^2 = (\frac{\omega'}{c})^2 \epsilon\mu_{0}, \hspace{3mm} k_{0}^2 = (\frac{\omega'}{c})^2 \epsilon_{0}\mu_{0}  \nonumber
\end{eqnarray}
in which\\
$
\left\{
  \begin{array}{ll}
    E^i, & \hbox{Incident field;} \\
    E^s, & \hbox{Scattered field;} \\
    G, & \hbox{Green's function;} \\
    j, & \hbox{Polarization current.}
  \end{array}
\right.
$\\
Eqns. (\ref{eq:OUT1}) describe the OUT with unknown dielectric characteristics $\epsilon$ which is illuminated from the circular antenna array; the scattered field is received by the receiving antennas on the same antenna array. As the ill-posed problem for the inverse system of determining $\epsilon$ in Eqns. (\ref{eq:OUT1}), approximation methods should be chosen. In \cite{article:Serguei Y. Semenov et al} a modified Rytov's approximation is used. Born approximation is also used for the above inverse problem concerning the objects with high contrast of $\epsilon$. In this case the Rytov's approximation gives better results. The algorithm in \cite{article:Serguei Y. Semenov et al} gives an accurate solution of the inverse problem in two-dimensional cases including image reconstruction of a phantom consisted of a semisoft gel cylinder. The gel phantom is immersed into the working chamber after being cooled in a refrigerator. It is shown that the dielectric situation inside the working chamber are affected by the temperature gradients. In addition, the dielectric properties of the phantom are also affected by non-isothermic conditions in the working chamber. Assuming that the frequency range from $2$ to $8$ GHz gives the most suitable results for microwave imaging \cite{article:J.C.Lin}, there are technical difficulties in building a tomographic system for the whole body concerning the frequency range. One of the reasons is that the acquisition time would be unrealistically long. However, at the lower frequency of about $0.9$ GHz suitable spatial resolution is achieved. In summary, the multifrequency range from $0.9$ to $3$ GHz is optimal for microwave tomographic imaging \cite{article:Serguei Y. Semenov et al}.

In \cite{article:J.CH.Bolomey}, a suitable method for quasi real-time microwave tomography for biomedical applications is presented. By simulating a focusing system characterized by small field depth and a variable focal length, a tomographic process is achieved in this work. The organ under test, which constitutes the scatterer, transforms the divergent wavefront from the focusing system into a convergent wavefront. An image, corresponding to a thin organ slice, from the divergent wavefront can be derived. By changing the focal length, different slices can be obtained resulting into a cross-section of the organ. From the measured field distribution, the slice images are deduced. Letting $d$ and $D$ be the length of the organ and the distance between the observation line and the slice, respectively, the length of the observation domain will be $2D + d$. The equivalent currents $\mathbf{J}$, responsible for the scattered field is \cite{article:J.CH.Bolomey}
\begin{equation} \label{eq:J_scatt}
  \mathbf{J}(x,y) = [K^2 (x,y) - K_{m}^2]\mathbf{E}_{t}(x,y)
\end{equation}
where $\mathbf{E}_{t}(x,y)$ and $K(x,y)$ are the total field and the wavenumber inside the organ, respectively; $K_{m}$ is the wavenumber of the homogeneous surrounding medium. For cylindrical objects, illuminated by a plane wave, the scattered field $\mathbf{E}_S$ is determined by \cite{article:J.CH.Bolomey}
\begin{equation} \label{eq:E_scatt}
  \mathbf{E}_{S}(x,y) = \int_{S}\mathbf{J}(x,y)H_{0}^{(2)} \left( K_{m}\sqrt{(x-x')^2 + (y-y')^2}\right)dx' dy'
\end{equation}
in which $H_{0}^{(2)}$ is the Hankel function of order zero and of the second kind. For both two-dimensional and the three-dimensional cases, such algorithms can be used to reconstruct $\mathbf{J}$ from the scattered field $\mathbf{E}_S$. Here, the reconstructed current is the image which appears as the convolution between the point-spread function of the focusing system and the induced current distribution in the organ. The \emph{method of angular spectrum} may be used for reconstruction of the current distribution from the scattered field \cite{article:Ch.Pichot et al}\cite{article:J.J.Stammes}.
\subsection{Direct Methods in Biological Imaging}
For the direct electromagnetic formulation, a classical approach considering a 2D version of the problem may be used as an alternative. A 3D version of the problem would otherwise be to describe the field properties using the Maxwell's equations which leads to a heavy 3D vectorial problem. In the 2D formulation, the biological object under test is considered to be nonmagnetic with constant dielectric properties along its vertical axis. The whole strategy in this approach is to convert the electromagnetic scattering problem into a radiating problem in the free space and a, so called, 2D scalar Electrical Field Integral Equation (EFIE). The implicit time dependence of $e^{-j\omega t}$, with $\omega$ as the radial frequency is also introduced. The homogeneous,- and inhomogeneous wave equations in this context are \cite{Tommy Henriksson:PhD}
\begin{equation} \label{eq:homogWE}
  (\nabla^2 + k_{1}^2) e^i (\vec{r}) = 0
\end{equation}
and
\begin{equation} \label{eq:inhomogWE}
  (\nabla^2 + k^2(\vec{r})) e(\vec{r}) = 0
\end{equation}
respectively. Here, $e^i(\vec{r})$, the incident field, is the propagation of a TM-polarized, single-frequency, time-harmonic electromagnetic wave and $e(\vec{r})$ is the total electric field; the constant wavenumber $k_{1}$ inside the homogeneous media, and the wavenumber $k$ are respectively as
\begin{equation} \label{eq:k1}
  k_{1}=\omega \sqrt{\mu_{0}\epsilon_{1}^*}
\end{equation}
and
\begin{equation} \label{eq:k}
  k^2(\vec{r})=\omega^2\mu_{0}\epsilon^*(r)
\end{equation}
where $\epsilon_{1}^*$ is the complex permittivity inside the homogeneous media, and $\epsilon^*(r)$ the complex permittivity of the inhomogeneous region. The total field, $e(\vec{r})$, as a superposition of the incident field and the scattered field $e^s(\vec{r})$ can be written as
\begin{equation} \label{eq:totField}
  e(\vec{r}) = e^i(\vec{r}) + e^s(\vec{r}).
\end{equation}
Introducing a new constant $C(\vec{r})$ as
\begin{equation} \label{eq:ConstantC}
  C(\vec{r}) = k^2(\vec{r}) + k_{1}^2
\end{equation}
together with the above equations will result into the following wave equation
\begin{equation} \label{eq:newWE}
  (\nabla^2 + k_{1}^2)e^s(\vec{r})= -C(\vec{r})e(\vec{r}).
\end{equation}
Associated with the scattered field $e^s(\vec{r})$ in Eqn. (\ref{eq:newWE}), an equivalent current $J(\vec{r})$ can be defined as
\begin{equation} \label{eq:currentJ}
  J(\vec{r})= C(\vec{r})e(\vec{r}).
\end{equation}
In fact, this equivalent current produces the scattered field and the wave equation above can now be written as \cite{Tommy Henriksson:PhD}
\begin{equation} \label{eq:ultimWE}
  (\nabla^2 + k_{1}^2)e^s(\vec{r})= -J(\vec{r}).
\end{equation}
A Green's function formulation for the inhomogeneous wave equation in (\ref{eq:ultimWE}) can be deduced to solve $e^s(\vec{r})$, that is
\begin{equation} \label{eq:GreenWE}
  (\nabla^2 + k_{1}^2)G(\vec{r},\vec{r'})= -\delta(\vec{r}-\vec{r'})
\end{equation}
where $\delta(\vec{r}-\vec{r'})$ is the Dirac delta function; the associated Green's function is
\begin{equation} \label{eq:GreenItself}
  G(\vec{r},\vec{r'}) = \frac{j}{H_{0}^{(1)}}(k_{1}|\vec{r}-\vec{r'}|)
\end{equation}
where $H_{0}^{(1)}$ is, as previously mentioned, the zero-order Hankel function of the first kind. By the aid of the Green's function formulation above, and the principle of superposition, the scattering field can be solved by
\begin{equation} \label{eq:GreenScatteringEq}
  e^s(\vec{r})=\int\int_{S}G(\vec{r},\vec{r'})C(\vec{r'})e(\vec{r'})d\vec{r'}.
\end{equation}
Considering (\ref{eq:totField}) and (\ref{eq:GreenScatteringEq}), the total field is finally expressed as the following integral formulation \cite{Tommy Henriksson:PhD}:
\begin{equation} \label{eq:GreenScTot}
  e(\vec{r})=e^i(\vec{r})+\int\int_{S}G(\vec{r},\vec{r'})C(\vec{r'})e(\vec{r'})d\vec{r'}.
\end{equation}
As the complex permittivity is known and the incident field $e^i(\vec{r})$ is given, the scattered field $e^s(\vec{r})$ will be computed as the direct formulation of the electromagnetic scattering problem. In such context, Eqns. (\ref{eq:GreenScatteringEq}) and (\ref{eq:GreenScTot}) can be solved, for example, by Moment Methods (MoM), see previous chapters. By this numerical method, two different two-dimensional configurations, by planar,-or cylindrical situated dipoles, are solved in \cite{Tommy Henriksson:PhD}. By assuming constant fields and dielectric properties in a rectangular cell as the OUT, the incident,- and the scattered field will be discretized as
\begin{equation} \label{eq:incidFieldDiscret}
  e^i(\vec{r_{n}})=\sum_{j=1}^N[\delta_{nj}-G(\vec{r_{n}},\vec{r_{j}})C(\vec{r_{j}})]e(\vec{r_{j}}), \hspace{3mm} n=1,2,...,N
\end{equation}
and
\begin{equation} \label{eq:scattFieldDiscret}
  e^s(\vec{r_{m}})=\sum_{j=1}^M[\delta_{nj}-G(\vec{r_{m}},\vec{r_{j}})C(\vec{r_{j}})]e(\vec{r_{j}}), \hspace{3mm} m=1,2,...,M
\end{equation}
where the region, i.e. the OUT, is discretized into $N$ cells and also $M$ receiving points for the observed scattered field; the Green's function $G$ can be computed analytically as depicted in \cite{Devaney:paper}. Numerical solution of this direct scattering problem will be used for creating image reconstruction algorithms for the inverse problem by which the unknown permittivity contrast distribution of the OUT will be found. Concerning biological image reconstruction by microwave methods, there are different approaches which are generally based on either \emph{radar techniques} or \emph{tomographic formulation} \cite{Hagnes et al:paper1}\cite{Hagnes et al:paper2}\cite{Hagnes et al:paper3}.
%\subsubsection{Rytov Approximation for Determining Permittivity}
%\subsubsection{Born Approximation for Determining Permittivity}

%\bibitem{article:J.C.Lin} "Frequency optimization for microwave imaging of biological tissues" \emph{in Proc. IEEE,} vol. 73, no. 2, pp. 374-375, 1985.
%%%%%% Chapter 4 %%%%%%%
\newpage
\appendix
\section{\emph{Appendix} \\\\Definitions }\label{App:AppendixA}
\begin{description}
  \item[Definition 1] A sequence of elements $\{v_{i}\}_{i=1}^\infty$ in a normed vector space $V$ is called a Cauchy (Fundamental) sequence if $\forall \epsilon > 0$ there exists $N\in\mathbb{N}$ such that $\|v_{i}-v_{j}\|<\epsilon$ for every $i,j>N$.
  \item[Definition 2] A sequence of elements $\{v_{j}\}_{j=1}^\infty$ in a normed vector space $V$ converges to an element $v$ if $v\in V$ and $\forall \epsilon > 0$ there exists $N\in\mathbb{N}$ such that $\|v_{j}-v\|<\epsilon$ for every $j>N$.
  \item[Definition 3] A normed space $V$ is complete (also called a Banach space) if every Cauchy (Fundamental) sequence in $V$ converges to an element in $V$.
  \item[Definition 4] Let $f:\mathbb{R}\longmapsto \mathbb{R}$. Then $f$ has compact support if $f(x)=0$ for every $x\in K$. That is $f(\mathbb{R}\setminus K)=0$ for some compact set $K\in \mathbb{R}$.
  \item[Definition 5] If $V$ is a linear space with a scalar product $<.,.>$ with a corresponding norm $\|\cdot \|$, then $V$ is said to be a Hilbert space if $V$ is complete, i.e., if every Cauchy sequence with respect to $\|\cdot \|$ is convergent.
\end{description}
%If $I=(a,b)$, then the space of square integrable functions on $I$ is defined as
%%%%%%%%%%%%%
%%%%%%%%%%%%%
\newpage
\section{\emph{Appendix} \\\\Square Integrable Functions}\label{App:AppendixB}
If $I=(a,b)$, then the space of square integrable functions on $I$ is defined as
\begin{equation}\label{hilbert}\nonumber
  L_{2}(I)=\{v: v  \hspace{2mm} is \hspace{2mm} defined \hspace{2mm} on \hspace{2mm} I \hspace{2mm} and \int_{I} |v|^2 dx < \infty \}
%L_{2}(I)=\{v: v  \mbox{\hspace{0.1mm} is \hspace{0.1mm} defined \hspace{0.1mm} on \hspace{0.1mm} I \hspace{0.1mm} and} \int_{I} v^2 dx < \infty \}
\end{equation}
%
%such that $v$ is defined on the interval \hspace \mbox{is} \hspace \mbox{defined} \hspace \mbox{on} \hspace I \hspace \mbox{and}
By defining a scalar product as
\begin{equation}\label{hilbert2}\nonumber
    (v,w)= \int_{I}vw dx
\end{equation}
and a corresponding norm ($L_{2}$ norm) as
\begin{equation}\label{hilbert3}\nonumber
    \|v\|_{L_{2}(I)}=\left[\int_{I}|v|^2 dx\right]^{1/2}
\end{equation}
The scalar product is such that
\begin{equation}\label{hilbert4}\nonumber
    |(v,w)| \leq \|v\|_{L_{2}(I)} \|w\|_{L_{2}(I)}
\end{equation}
which means that the above integral exists if $v$ and $w \in L_{2}(I)$.
\newpage
%\end{thebibliography}
\end{document}